\documentclass[pdflatex,sn-mathphys-num]{sn-jnl}

\usepackage{graphicx}%
\usepackage{multirow}%
\usepackage{amsmath,amssymb,amsfonts}%
\usepackage{amsthm}%
\usepackage{mathrsfs}%
\usepackage[title]{appendix}%
\usepackage{xcolor}%
\usepackage{textcomp}%
\usepackage{manyfoot}%
\usepackage{booktabs}%
\usepackage{algorithm}%
\usepackage{algorithmicx}%
\usepackage{algpseudocode}%
\usepackage{listings}%
\usepackage{hyperref}%
\usepackage{cleveref}%
\usepackage{siunitx}%
\usepackage{longtable}

\theoremstyle{thmstyleone}%
\theoremstyle{thmstyletwo}%

\theoremstyle{thmstylethree}%

\begin{document}

\title[Article Title]{Balancing Cost Savings and Import Dependence in Germany's Industry Transformation}


\author*[1]{\fnm{Toni} \sur{Seibold}}\email{seibold@tu-berlin.de}
\author[1]{\fnm{Fabian} \sur{Neumann}}\email{f.neumann@tu-berlin.de}
\author[2]{\fnm{Falko} \sur{Ueckerdt}}\email{ueckerdt@pik-potsdam.de}
\author[1]{\fnm{Tom} \sur{Brown}}\email{t.brown@tu-berlin.de}

\affil*[1]{\orgdiv{Digital Transformation in Energy Systems}, \orgname{Technische Universität Berlin}, \orgaddress{\street{Einsteinufer 25}, \city{Berlin}, \postcode{10587}, \state{State}, \country{Germany}}}

\affil[2]{\orgdiv{Research Department 3—Transformation Pathways}, \orgname{Potsdam Institute for Climate Impact Research}, \orgaddress{\street{P.O. Box 60 12 03}, \city{Potsdam}, \postcode{14412}, \country{Germany}}}

\abstract{
Greenhouse gas emissions from the steel, fertiliser and plastic industries can be mitigated by producing their precursors with green hydrogen.
In Germany, green production may be economically unviable due to high energy costs.
This study quantifies the 'renewables pull' of cheaper production abroad and highlights trade-offs between cost savings and import dependence.
Using a detailed European energy system model coupled to global supply curves for hydrogen and industry precursors (hot briquetted iron, ammonia and methanol), we assess five scenarios with increasing degrees of freedom with respect to imports.
We find that precursor import is preferred over hydrogen import because there are significant savings in hydrogen infrastructure.
Cost savings in the German industry sector from shifting precursor production to European partners compared to domestic production are at \SI{4.1}{bnEUR/a} or \SI{11.2}{\percent}.
This strategy captures \SI{47.7}{\percent} of the cost savings achievable by precursor import from non-European countries, which lowers industry costs by \SI{8.6}{bnEUR/a} (\SI{23.3}{\percent}).
Moving energy-intensive precursor production abroad allows Germany to save costs while still retaining a substantial share of subsequent value-creating industry.
However, cost savings must be weighed against the risks of import dependence, which can be mitigated by sourcing exclusively from regional partners.
}



\maketitle
Improving economic competitiveness has joined climate protection as one of the leading policy objectives in the European Union, as recently set out in the Draghi report \cite{european_commission_european_political_strategy_centre_future_2025}.
Some of the key priorities are lowering energy prices, raising resilience and diversifying energy supply.

For Germany, addressing these priorities is especially challenging for three main reasons.
First, the country is currently heavily dependent on fossil fuel imports.
In 2021, Germany imported around \SI[group-separator={,}, group-minimum-digits=4]{2930}{PJ} of natural gas and \SI[group-separator={,}, group-minimum-digits=4]{3455}{PJ} of crude oil which together make up \SI{51.5}{\percent} of its primary energy demand \cite{eurostat_complete_2022}.
Since domestic renewable energy yields are comparatively low and space is limited, even a climate neutral energy system may have to depend on energy import.
Second, access to cheap fossil fuel imports in the past helped build a strong energy-intensive industry which now accounts for about \SI{20}{\percent} of Germany's gross value added \cite{statistisches_bundesamt_bruttowertschopfung_2025}.
With some of the highest electricity prices in the world, it is becoming harder to keep energy-intensive industries competitive and prevent them from moving abroad.
Third, Germany has set itself a particularly ambitious target: climate neutrality by 2045, which is five years ahead of the EU.
This increases the pressure to find solutions that balance environmental and economic needs.

The industry sector is responsible for \SI{18.2}{\percent} of greenhouse gas emissions in Germany \cite{umweltbundesamt_ubersicht_2023}.
Currently, one solution posited in political discussion for processes that are not easy to electrify is green hydrogen, either produced domestically or imported from abroad \cite{federal_ministry_for_economic_affairs_and_climate_action_bmwk_nationale_2020, federal_ministry_for_economic_affairs_and_climate_action_bmwk_fortschreibung_2023}.
Hydrogen would allow energy-intensive processes to switch from fossil production routes, e.g. moving steel making from coke-based blast furnaces to direct reduced iron (DRI) plants \cite{dpa_stahlunternehmen_2025}.
To make hydrogen available to industrial sites, a national strategy has been put in place to roll out a hydrogen network \cite{federal_ministry_for_economic_affairs_and_climate_action_bmwk_nationale_2020, federal_ministry_for_economic_affairs_and_climate_action_bmwk_fortschreibung_2023}.
Studies show that hydrogen can play a pivotal role in the future energy system and a hydrogen pipeline network would be beneficial for Europe and Germany in particular \cite{lux_role_2022, schob_role_2023,luderer_energiewende_2025,sensfus_langfristszenarien_2025}.
At the same time, concerns have been raised that large-scale imports from non-European countries could create new dependencies, whereas domestic self-sufficiency would be feasible only at higher costs \cite{nunez-jimenez_competitive_2022}.

Since hydrogen transport by ship is associated with high losses (e.g. for liquefaction or hydrogenation and dehydrogenation of liquid organic hydrogen carriers) and transport across the continent requires pipeline build out, there is a growing discussion on splitting value chains \cite{hampp_import_2023}.
This refers to the import of hydrogen derivatives that serve as precursors for many materials in energy-intensive industry, such as hot briquetted iron (HBI) from a DRI process for steel production, ammonia for fertilisers and methanol for high value chemicals (HVC).
The final processing of the materials, e.g. of HBI in an electric arc furnace to steel, upgrading ammonia to urea, or processing methanol into olefins and aromatics, could still take place in Germany.
Multiple studies have evaluated green supply chains for ammonia \cite{egerer_economics_2023, kim_green_2024}, methanol \cite{fasihi_global_2024, sollai_renewable_2023} and steel \cite{lopez_towards_2023, ellersdorfer_unlocking_2024, bilici_global_2024} as well as all of the above \cite{hampp_import_2023}.
One study has incorporated the import of these energy carriers into a comprehensive European energy system model to investigate the impact on hydrogen infrastructure needs \cite{neumann_green_2025}.
Cost benefits were found by outsourcing the production of energy intensive precursors to non-European countries with high renewable potential.
This mechanism, introduced in \cite{samadi_renewables_2023} as the 'renewables pull', describes how countries with abundant renewable resources attract energy-intensive processes.
As production shifts away from fossil fuels, electricity prices become the dominant driver of production costs.
This approach would significantly reduce primary energy demand in Germany and Europe, thereby alleviating pressure on the European energy system.
However, in addition to lower electricity prices in non-European countries which tend to attract energy-intensive industry sectors, policies and measures also exist to protect current industrial locations \cite{nykvist_renewables_2025}.
Moving solely the production of precursors abroad offers a trade-off between outsourcing industrial processes and retaining value creation within Europe.

Verpoort et al. investigated this trade-off for Germany by assuming different electricity price spreads and found that a \SI{40}{EUR/MWh} difference leads to cost savings of \SI{38}{\percent} while keeping a significant share of value creation in Germany \cite{verpoort_impact_2024}.

Up until now research has focused on global or European import infrastructure, or looked at national consequences in very simplified settings.
In this paper, we bridge this gap in the literature and focus on the implications for one particular country with a large share of energy-intensive industries, Germany, but embedded in a detailed European energy system model with global supply chains for green materials.
This approach eliminates the need to assume exogenous price differences, since they are modelled endogenously and allows us to distinguish between production in Germany, Europe and import from non-European countries.
Our model enables us to address the following questions:
\begin{itemize}
    \item What is needed to keep the production of green industrial precursors in Germany competitive with imports?
    \item Would outsourcing precursor production to the rest of Europe represent a compromise between maintaining competitiveness and energy security compared to importing from a global market?
    \item How would the relocation of precursor production propagate through the German energy system, particularly with respect to energy prices and infrastructure needs?
\end{itemize}

To explore these questions we use an open source energy system model, \mbox{\textit{PyPSA-DE}}, that represents Germany embedded in Europe \cite{PyPSA-DE}.
We adapt this model with the global import possibilities from Neumann et al. \cite{neumann_green_2025}.
We impose climate neutrality corresponding to the German 2045 and EU's 2050 target and then successively relax constraints on energy-intensive imports in several scenarios.

All scenarios assume that industrial precursors are produced via hydrogen-based supply chains.
The model includes three hydrogen production pathways: green hydrogen (via electrolysis powered by renewable electricity), blue hydrogen (via steam methane reforming (SMR) with carbon capture (CC)) and grey hydrogen (SMR without CC).
In addition to the methanolisation of hydrogen and CO\textsubscript{2}, a biomass-based production route is considered.
Hot briquetted iron (HBI) is produced exclusively via a hydrogen DRI process.
The final step in steel production is carried out in an electric arc furnace (EAF), which remain in the model at their current locations to account for established ties to the manufacturing industry.
Ammonia is produced using a hydrogen-fed Haber-Bosch process, with nitrogen supplied via an air separation unit that requires electricity.
Methanol serves both as a final product in industry and as feedstock for high-value chemicals through the methanol-to-olefins/aromatics conversion route.
At present, feedstock for these chemicals is fossil oil-based, produced through steam cracking into shorter hydrocarbons.
Because the total efficiency of Fischer-Tropsch fuels following this production route is not competitive with the methanol-based pathway, this option is excluded from our model.
The precursor production pathways are illustrated in Fig.~\ref{fig:industry_endogenized} and a detailed description of the model's scope is provided in the \nameref{sec:methods} Section.

\begin{figure}[htbp]
    \centering
    \includegraphics[width=0.9\textwidth]{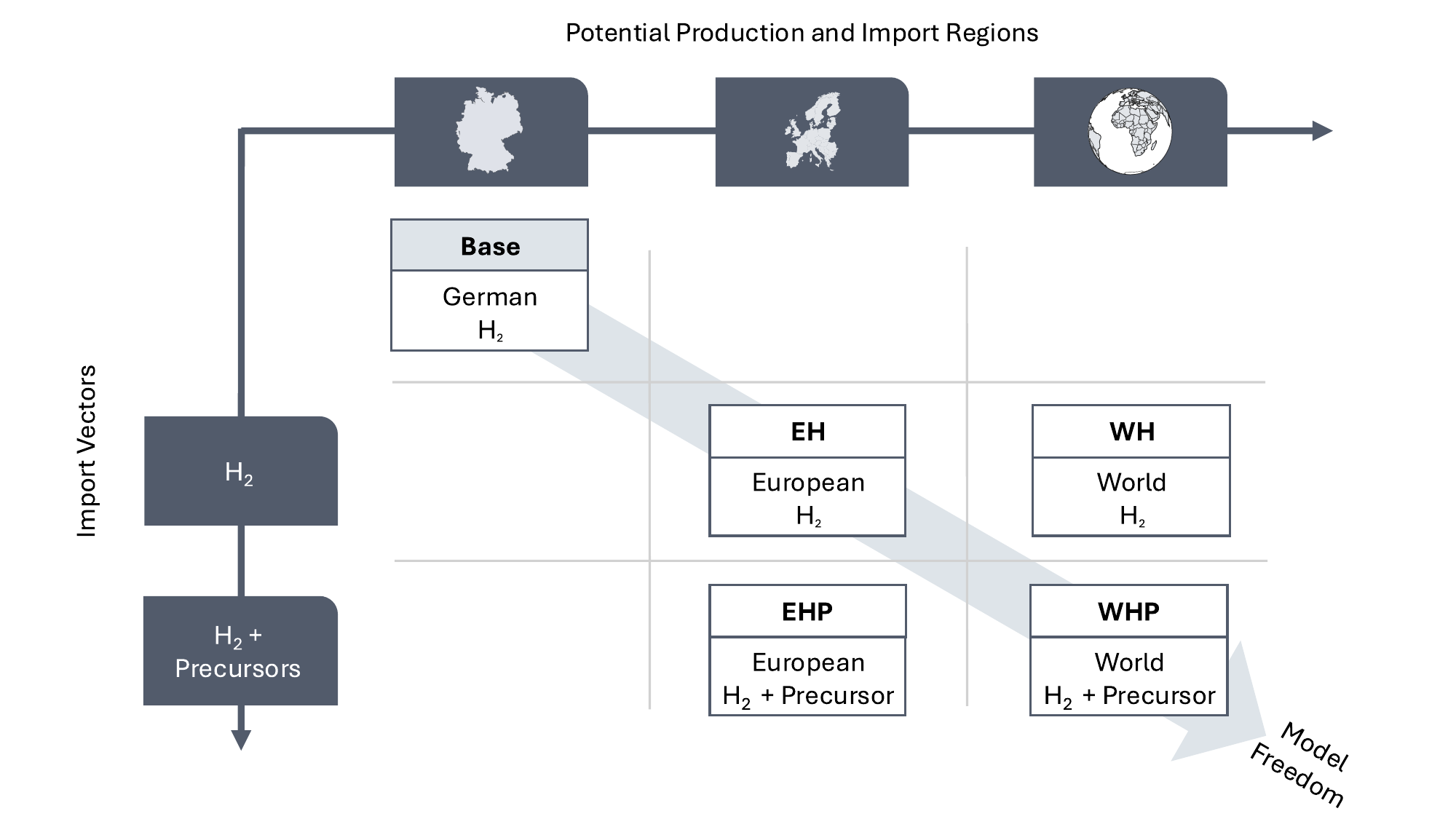}
    \caption{Scenario framework to explore the influence of imports on the German industry and energy system. While the Base scenario represents a domestic industry with hydrogen self-sufficiency, 
    the geographic scope of allowed imports expands on the horizontal axis, while the diversity of import products increases on the vertical axis.}\label{fig:scenario_explanation}
\end{figure}

Fig.~\ref{fig:scenario_explanation} provides an overview of the scenario design.
The Base scenario is used as a benchmark to compare different industrial development pathways.
Industry sites are sustained by producing the industry precursors HBI, ammonia and methanol domestically.
This is the case not only for Germany, which our study focuses on, but all European countries.
For Germany, the hydrogen demand must be met from domestic sources and trade is forbidden.
While Germany is modelled with 30 nodes, the neighboring countries are only represented by one node each.
Due to their low geographical resolution, which prevents internal grid bottlenecks from being represented, the annual net electricity import to Germany is limited to a volume of \SI{125}{TWh} (more than double the import volume in 2022 \cite{bundesnetzagentur_marktdaten_nodate}).
Since this study puts a focus on the industry sector, transport fuels (methanol for shipping and Fischer-Tropsch fuels for aviation and the agriculture sector) can be imported from non-European countries across all scenarios to cover German and European demand, reflecting today's import patterns for fossil fuels with green substitutes.

Compared to the Base scenario, the scenarios EH (\textbf{E}urope + \textbf{H}ydrogen) allows the import of hydrogen from European partners by a pipeline network.
Scenario WH (\textbf{W}orld + \textbf{H}ydrogen) relaxes the model further by additionally allowing hydrogen imports from non-European partners via pipeline and ship.
This global import is available not only to Germany but throughout Europe.
Industrial sites are kept at their current location.

Scenarios EHP (\textbf{E}urope + \textbf{H}ydrogen + \textbf{P}recursors) and WHP (\textbf{W}orld + \textbf{H}ydrogen + \textbf{P}recursors) relax the model further by allowing energy intensive precursors to be produced in European and non-European countries respectively.
This allows Germany to import HBI, ammonia and methanol as industry precursors from (non-)European partners.
The industrial plants for the further processing of the precursors to steel, fertiliser and HVC remain in their current locations.

The optimization problem's solution space is expanded from the Base to the WHP scenario and the model can choose endogenously how much is imported.

\section*{Results}\label{sec2}
In comparison to the Base scenario, the total system cost in Germany decreases in all scenarios.
By importing, Germany avoids domestic investments in renewable energies and power-to-X plants.
To measure the economic benefits of relocating hydrogen and precursor production, we account for the cost of supplying Germany's demand, regardless of where the production is located.

Our model includes exogenous and endogenous demand profiles for electricity, hydrogen, heat, biomass, liquid hydrocarbons, methanol, HVC, ammonia and steel.
From the optimization results, we extract the Langragian multipliers $\lambda_{i,t}$, which represent the marginal price for an energy carrier or feedstock in region $i$ and time step $t$.
When multiplied by the corresponding consumption $d$, we obtain the consumer costs $c_{cons}$:

\begin{equation}
    c_{cons}  = \sum_{i \in I_{DE}} \sum_{t \in T} \lambda_{i,t} \cdot d_{i,t}
\end{equation}

\noindent Appendix~\ref{A1} provides further details on the model equations.

\noindent\textbf{Allowing precursor import reduces industry cost up to 23.3 \%}

\begin{figure}[htbp]
    \centering
    \includegraphics[width=0.95\textwidth]{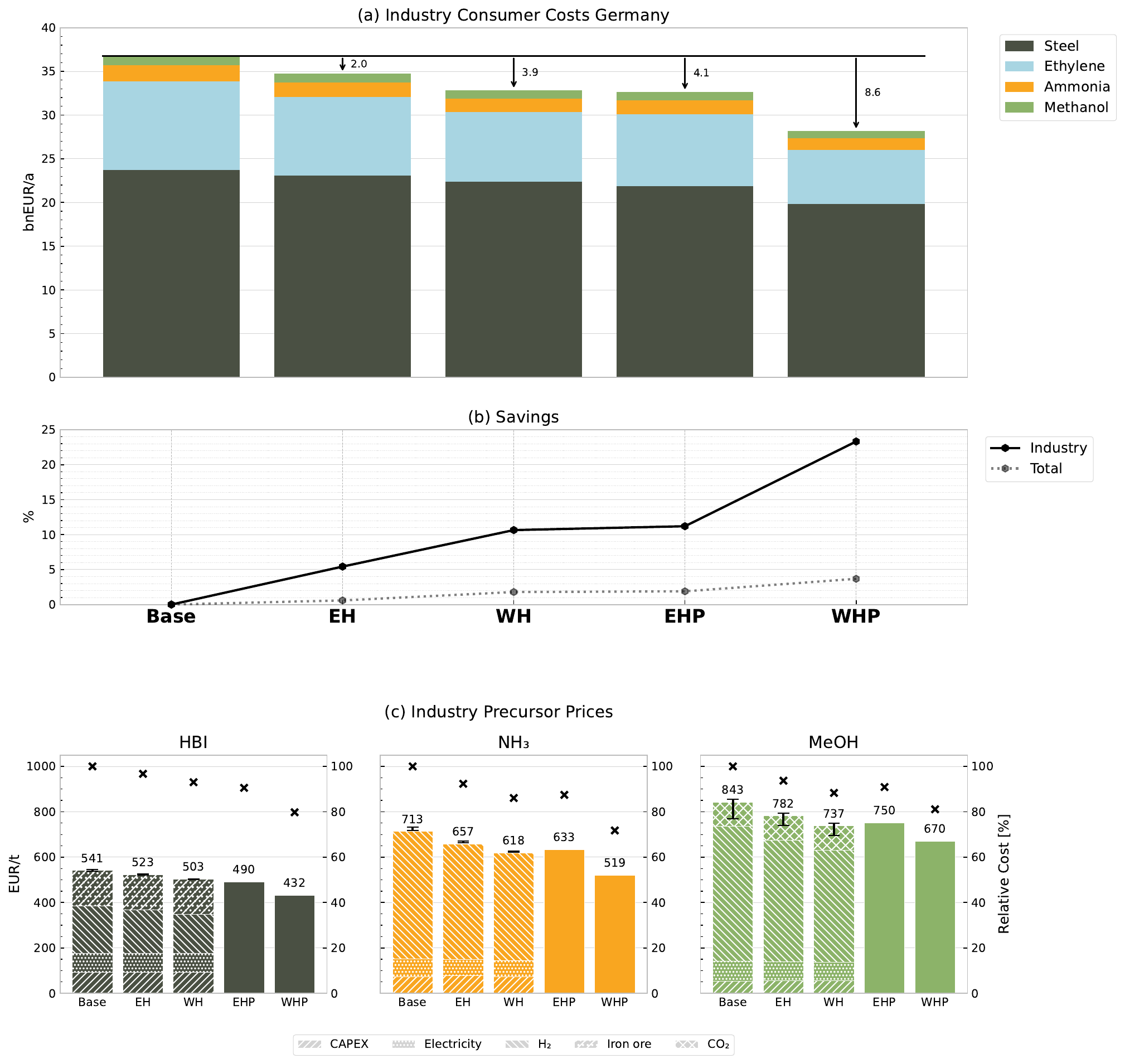}
    \caption{Consumer costs to meet industrial demand in Germany with savings across different levels of independence (a), relative cost savings overall and the industrial sector compared to the Base scenario (b) and prices of industry precursors in Germany (c).}\label{fig:consumer_costs}
\end{figure}

\noindent As depicted in Fig.~\ref{fig:consumer_costs}, allowing the import of industry precursors from European or non-European countries provides substantial economic benefits.
Industry consumer costs amount to \SI{36.8}{bnEUR/a} in the Base scenario.
European hydrogen imports (EH) yield savings of \SI{2.0}{bnEUR/a} (\SI{5.4}{\percent}), allowing hydrogen import from non-European partners (WH) reduces costs by \SI{3.9}{bnEUR/a} (\SI{10.7}{\percent}) and European precursor imports (EHP) achieve \SI{4.1}{bnEUR/a} (\SI{11.2}{\percent}).
Giving the system full flexibility to import both hydrogen and precursors from non-European partners (WHP) decreases industry consumer costs further, by \SI{8.6}{bnEUR/a} (\SI{-23.3}{\percent}) down to \SI{28.2}{bnEUR/a}.
In comparison, relocation limited to European partners (EHP) delivers cost savings corresponding to \SI{47.7}{\percent} of the maximum WHP benefit.

In terms of industry consumer cost savings, the WH and EHP scenarios yield similar results, though for different reasons.
In the EHP scenario, relocating precursor production enables the use of Europe's most favorable renewable resources.
Hydrogen and electricity demand for precursors is relocated to countries with better resources.
By placing renewables, electrolysis, and precursor production in close proximity, the system avoids major grid and pipeline expansion.
Hydrogen is supplied within Europe, as in the EH scenario, but in a more efficient configuration.
In the WH scenario, cost savings arise from a different strategy: combining domestic production with imports from non-European countries.
Europe exploits its most favorable renewable sites, producing \SI{504.2}{TWh/a} of hydrogen locally, while additional imports from outside Europe allow demand to be met without developing renewables in less suitable locations.

The cost reductions for industrial precursors are substantial: compared to the Base scenario, wholesale prices for HBI, ammonia and methanol decline by up to \SI{-20.2}{\percent}, \SI{-28.2}{\percent} and \SI{-21.0}{\percent} respectively.
The largest cost decrease is observed for ammonia, reflecting the high contribution of hydrogen and electricity to its overall cost.
In contrast, HBI is also influenced by costs for iron ore, whereas methanol depends on CO\textsubscript{2}.
Not only the industrial but also total consumer costs are depending on the availability of imports.
Fig.~C.4 shows the total consumer costs in Germany and absolute savings.

\noindent \textbf{Precursor import preferred over hydrogen import}

\begin{figure}[htbp]
    \centering
    \includegraphics[width=0.95\textwidth]{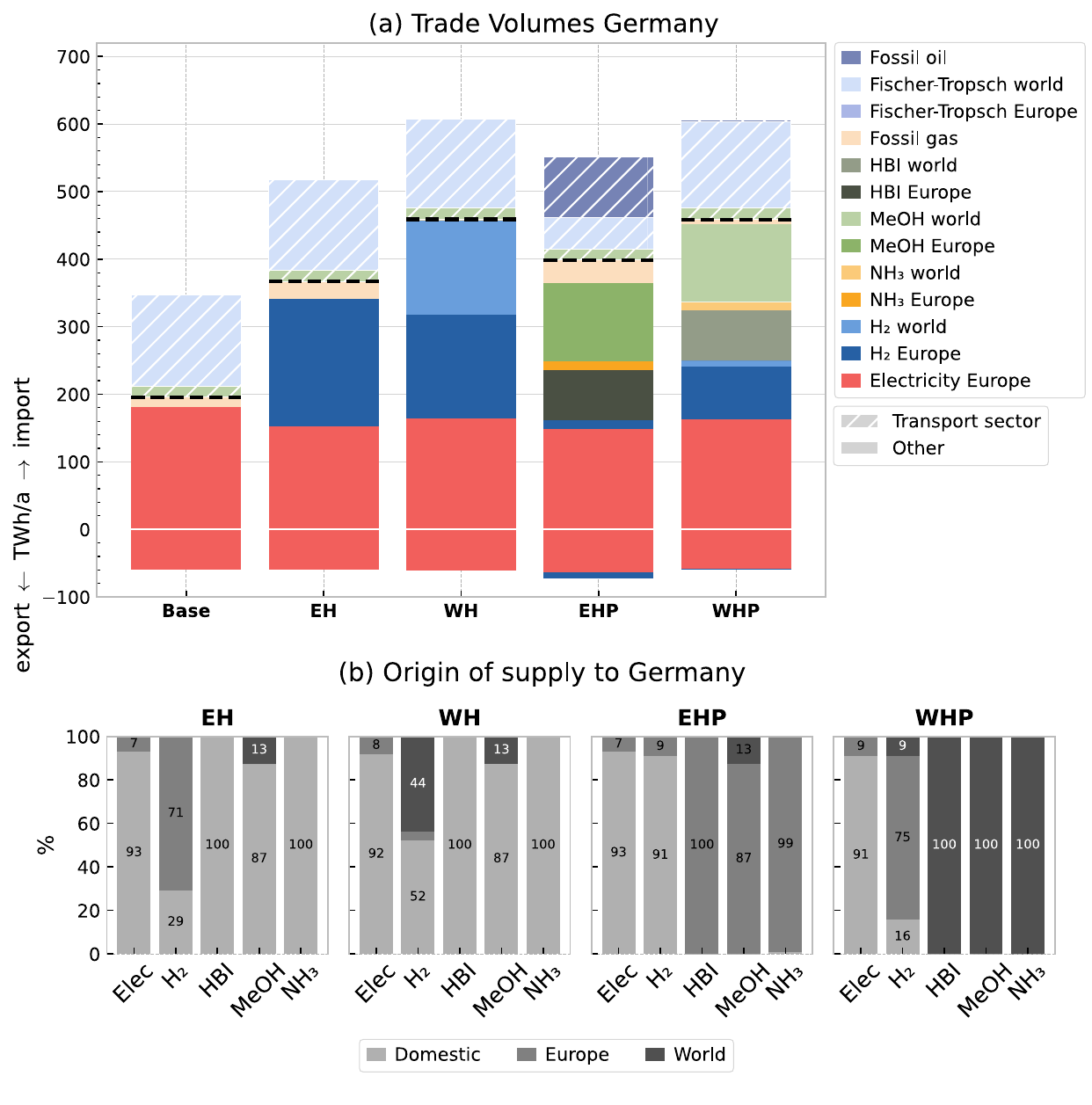}
    \caption{German annual import of energy carriers across scenarios (a) and the share they contribute in meeting the total demand within the country (b). HBI volume is multiplied by 2.1 which is the hydrogen in MWh needed to reduce one ton of iron ore. Since the transport sector is not explicit part of this analysis, fuels are shown in the upper part of each trade balance.}\label{fig:import_balance}
\end{figure}

\noindent Fig.~\ref{fig:import_balance} shows the import balances for each scenario and the origin of Germany's electricity, hydrogen, HBI, ammonia and methanol.

In the Base scenario there are no precursor or hydrogen imports, since hydrogen has to be supplied exclusively from domestic electrolysis.
The net electricity import limit is fully exploited with \SI{125}{TWh/a}.

In contrast, scenarios EH and WH show annual hydrogen imports of \SI{188.1}{TWh} in EH and \SI{239.2}{TWh} in WH  from European and non-European partners, reducing the need for domestic hydrogen production.
Nevertheless, domestic hydrogen demand remains high because precursors continue to be produced within Germany.
This reults in electricity imports of \SI{92.7}{TWh} and \SI{103.7}{TWh}, respectively.

In the EHP scenario, precursor production of methanol, ammonia, and HBI is relocated to European partners.
With the hydrogen demand for precursor production now outsourced, local production in Germany reduces to the amount that is economical to absorb surplus renewable electricity, reducing hydrogen imports to just \SI{4.7}{TWh}.
This hydrogen is used in the power sector for backup generation.
The share of hydrogen that is domestic actually rises, since the overall consumption declines rapidly, while the cheap local hydrogen from surplus renewables stays roughly constant.

In the WHP scenario, all precursor production is outsourced to non-European partners.
The import of hydrogen of \SI{85.5}{TWh} allows Germany to become almost completely independent from fossil fuels.
Hydrogen mostly covers the demand in times of low renewable availability with dispatchable capacities.
When given the choice, the model prefers non-European precursor imports over direct hydrogen imports.

Some fossil gas and oil import is allowed in the scenarios, since their emissions can be offset by the sequestration of CO\textsubscript{2} from biomass.
In scenarios where hydrogen-based imports are limited to Europe, fossil gas is used primarily in backup generation (gas combined heat and power plant (CHP) and open cycle gas turbine (OCGT)) with and without carbon capture.
The model identifies this pathway as cost-effective, since building additional renewables in less favorable regions for hydrogen production is more expensive than combining fossil gas with carbon dioxide removal.
In contrast, this effect does not appear in the WH and WHP scenarios, where cheaper non-European hydrogen is available and displaces the role of unabated fossil gas usage with compensation by carbon dioxide removal (CDR).

The energy system model includes the option of producing synthetic natural gas (SNG) through methanation of hydrogen with CO\textsubscript{2}.
However, this green alternative remains economically uncompetitive when compared to biogas and fossil natural gas (\SI{22.9}{EUR/MWh}) combined with carbon capture and sequestration (CCS) even though biomass availability is restricted to residues and the carbon sequestration potential is \SI{200}{Mt/a} in Europe.
Consequently, natural gas continues to play a role, particularly in meeting exogenous industrial demand.
It is only used for dispatchable backup capacities when hydrogen is scarce.

Methanol for shipping is sourced from non-European partners throughout Europe in all scenarios.
In Germany this methanol share makes up \SI{13}{\percent} of total methanol demand as depicted in Fig. \ref{fig:import_balance} (b).
As with Germany, the entire European methanol demand for shipping is met through imports from non-European partners across all scenarios, totaling \SI{503}{TWh} as depicted in Fig.~D.13.

For meeting oil demand for aviation and in the agricultural sector, there are three competing possibilities:
the Fischer-Tropsch synthesis in non-European countries, the Fischer-Tropsch synthesis in Europe or by using fossil oil.
In scenarios Base, EH, WH and EHP between \SI{507.6}{TWh} and \SI{513.0}{TWh} of fossil oil and \SI{281.3}{TWh} to \SI{289.9}{TWh} non-European imports are used to meet demand.
Fischer-Tropsch synthesis outside Europe is economically more attractive than domestic production, despite relying solely on CO\textsubscript{2} from direct air capture (DAC).
Only the WHP scenario frees up enough capacities to produce \SI{94.6}{TWh} of Fischer-Tropsch fuel in Europe.

\noindent \textbf{Relocating precursor production lowers hydrogen prices and the need for infrastructure}

\noindent Using the best renewable resources while avoiding suboptimal sites leads to lower hydrogen prices.
By relocating precursor production abroad, demand for hydrogen and consequently electricity is reduced, allowing generation to be concentrated at the most favorable locations.
This not only lowers the cost of electricity and hydrogen but also reduces the need for extensive hydrogen infrastructure.
As precursors HBI, ammonia and methanol are imported, the domestic demand for hydrogen decreases, further reducing the need for hydrogen production and transport capacity.

In the EH scenario, hydrogen imports from European partners maintain high domestic demand but lower the marginal price to \SI{89.9}{EUR/MWh} (-\SI{9}{\percent}).
Here, Spain and Italy supply central Europe via pipeline connections.
Allowing imports from non-European partners in the WH scenario reduces the price further to \SI{80.99}{EUR/MWh} (-\SI{18.2}{\percent}), with the Italy-North Africa pipeline corridor becoming a key supply route.

In the EHP scenario, precursor production shifts to European countries with high solar irradiation (e.g. Spain) or high wind potential (e.g. Denmark and the UK), lowering the need for long-distance hydrogen transport and infrastructure expansion.
In the Base scenario a total of \SI{71.1}{TWkm} of pipelines are built in Europe which declines to \SI{35.0}{TWkm} in the EHP scenario.
In Spain, the model's maximum technical potential for solar PV is largely exhausted, which is why we examine allowing more land for PV deployment in a sensitivity analysis in Appendix~\ref{A4}.
As a result, these geographic constraints keep the hydrogen price high at \SI{86.51}{EUR/MWh}.
Finally, the WHP scenario combines global hydrogen and precursor imports, shifting production to regions with abundant renewable resources.
Instead of transporting hydrogen across Europe, semi-finished products are imported, substantially reducing infrastructure needs and yielding a more efficient overall system design.

\begin{figure}[htbp]
    \centering
    \includegraphics[width=0.8\textwidth]{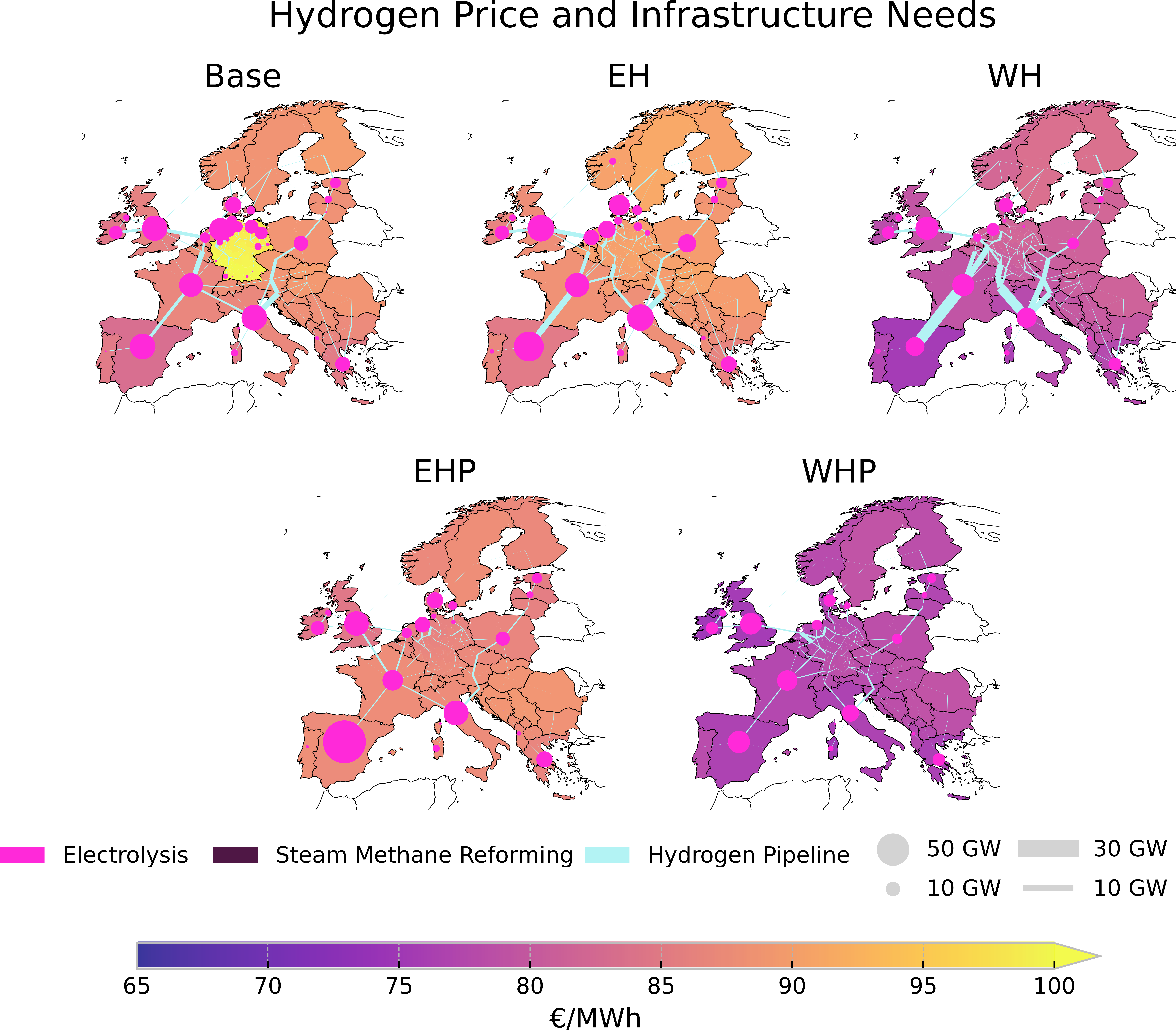}
    \caption{Hydrogen production, infrastructure and price across scenarios spatially resolved for each model region.}\label{fig:hydrogenmap}
\end{figure}

\noindent \textbf{Sensitivity of results to capital costs, electrolyser costs, sequestration potential and land use}

\noindent Parts of the results depend on how the model is parameterised.
In Appendix~\ref{A4} we explore some dependencies on the weighted average cost of capital (WACC) of the non-European imports, the investment costs of electrolysis units, the potential volume of CO\textsubscript{2} sequestration and the amount of land made available to solar.
Making more land available for solar by raising the density of PV from \SI{5.1}{MW/km\textsuperscript{2}} to \SI{25}{MW/km\textsuperscript{2}} leads to more cost savings predominantly in the Base, EH and EHP scenario where imports from non-European imports are limited.
The higher sequestration potential allows Europe and Germany to rely more on fossil fuels especially in the scenarios where the import from non-European partners is limited to transport fuels.
Assuming a low investment cost for electrolysis units of \SI{500}{EUR/kW} (compared to \SI{1000}{EUR/kW} in the main scenarios) leads to savings in the order of \SI{10}{bnEUR/a} in the overall consumer costs.
However for those sensitivities the overall conclusion remains the same: importing precursors from European countries reaches \SI{52.4}{\percent} to \SI{63.5}{\percent} of the savings achieved by importing from non-European partners.
Only increasing the WACC of global imports from \SI{7}{\percent}, the rate assumed for Europe, to \SI{10}{\percent}, raises the cost of non-European imports so that the EHP scenario attains \SI{78.3}{\percent} of the benefit of the WHP scenario.

\section*{Conclusion and Discussion}\label{sec:conclusion}

This study applies a sector-coupled energy system model to assess the effects of relocating hydrogen and industrial precursor production from Germany to European and non-European countries.
By endogenizing the production and trade of key energy carriers (hydrogen, HBI, ammonia and methanol) the model captures system-wide cost and infrastructure implications under climate policy.

Relocating industrial precursor production results in substantial cost savings.
Imports from non-European countries are the most cost-effective, reducing industrial consumer costs by \SI{8.6}{bnEUR/a} or \SI{23.3}{\percent}.
Restricting imports to Europe achieves \SI{47.7}{\percent} of those savings.
If policy-makers have to subsidise the difference to keep European precursors competitive in a world market, this would translate into \SI{4.5}{bnEUR/a} in subsidies for the German industry sector only.
Also only allowing the import of hydrogen from European and non-European partners while keeping all energy-intensive industry would lead to a gap of \SI{4.7}{bnEUR/a}.

Beyond direct cost savings in the industry sector, relocating precursor production also affects system design.
It lowers hydrogen prices and reduces the need for an extensive hydrogen pipeline infrastructure.
In scenarios allowing precursor imports, the model consistently favors importing precursors over hydrogen, reflecting not only the high transport costs and conversion losses that make hydrogen a less competitive import vector, but also the outsourcing of the substantial electricity demand associated with precursor production.

The lower domestic demand for hydrogen puts in question some of the extensive hydrogen network plans in Europe \cite{federal_ministry_for_economic_affairs_and_climate_action_bmwk_nationale_2020,federal_ministry_for_economic_affairs_and_climate_action_bmwk_fortschreibung_2023, slowinski_european_2023}.
If much of the industry demand falls away, there may not be enough demand for backup power generation to sustain large hydrogen infrastructure.
Hydrogen derivatives like methanol or ammonia, which as liquids are easier to store and transport, could then be used to supply this residual backup need \cite{glaum_minimal_2025}.

These findings highlight that the cost-optimal decarbonization of the industry sector has some clear trade-offs with the vulnerabilities of extended value chains.
While global imports provide the lowest costs, they may lead to increased dependence on external markets for core industrial inputs.
Restricting imports to Europe captures half of the global benefit (or up to \SI{78.3}{\percent} if the global WACC is higher), which offers a compromise between supply chain security and competitiveness.
Countries may also choose to pursue mixed strategies, keeping some domestic production alongside importing precursors from political partner countries, and allowing the remainder to be obtained from global markets.

Beyond geopolitical dangers such as the dependency on a small number of export countries, felt acutely in Europe during the 2022 gas crisis, or the threat of blockade, not experienced in Europe since the Second World War, compliance with regulation may be an additional reason to keep supply chains within Europe.
Europe has strict rules on the production of hydrogen-based renewable fuels of non-biological origin as well as for the sourcing of sustainable carbon \cite{zeyen_mitigating_2021}.
If direct air capture, which is assumed as the CO\textsubscript{2} source non-European imports, turns out to be scarce, then producers will have to rely on biomass for carbonaceous fuels.
Given repeated disputes surrounding the import of supposedly sustainable biomass to Europe, this may be an additional motivation to restrict supply chains to regions with compatible governance \cite{mdr_betrug_nodate,noauthor_climate_2019}.

Looking towards the full value chain, outsourcing precursor production represents a 'sweet spot' where only the energy-intensive part of the value chain, with low value added and a small number of jobs, is moved abroad \cite{verpoort_transformation_2024}.
This allows Germany to retain both the further processing of the precursors into steel, fertilisers and plastics, as well as the high value-added parts of the supply chain like automobile production and other manufacturer goods.
The final processing steps can then be done near the final consumers, tailored to their needs.
While the further erosion of value chains, whereby ever more of the value chains move abroad, is a potential danger, the cost savings from precursor import would allow more fiscal space to support these more valuable parts of the supply chains.

This paper provides orientation for addressing these important trade-offs.
Geopolitics, industrial policy, the energy transition and infrastructure planning are closely linked and need to be coordinated domestically and with international partners.
The 'sweet spots' highlighted here, such as partial splitting of supply chains at the precursor stage or allowing relocation to political partners, may help navigate these challenges while maintaining competitiveness as well as energy security.

\section*{Methods}\label{sec:methods}

\noindent\textit{European energy system model with a focus on Germany}

\noindent This analysis builds on the open-source energy system model \textit{PyPSA-DE} \cite{PyPSA-DE}, representing all EU member states together with Switzerland, Norway, the United Kingdom and the Balkan countries.
Most countries are modelled with one to three spatial nodes, while Germany is represented in greater detail with 30 nodes.
A visual representation of the model scope is depicted in Fig.~\ref{fig:hydrogenmap}.
The model accounts for energy and feedstock demand across five sectors: electricity, heating, transport, industry and agriculture.
Electricity demand is taken from ENTSO-E data via \textit{Open Power System Data}, with demand components already attributed to other modelled sectors subtracted to avoid double counting \cite{muehlenpfordt_time_2020}.
The heating sector includes space and water heating in residential buildings, assuming a retrofit of the existing building stock that significantly reduces demand \cite{zeyen_mitigating_2021}.
Heat can be provided by biomass and gas boilers and heat pumps.
In urban areas demand can also be met by gas and biomass CHP plants that are connected to district heating systems.
The transport sector covers light and heavy road transport, domestic and international shipping and aviation.
Road transport demand is distributed across battery electric, fuel cell and internal combustion engine vehicles according to an exogenously defined drivetrain technology mix.
Aviation demand is met by kerosene, either fossil-based or produced via Fischer-Tropsch synthesis or a methanol-to-kerosene route, while shipping demand is assumed to be fully covered by methanol-powered vessels.
In the industry sector, demand is divided between electricity, hydrogen, biomass and gas, following pre-defined fuel and process switches (e.g. methane for high-temperature heat, biomass for low-temperature heat) which can serve both as feedstock and energy carrier.
The agricultural sector adds additional demand for electricity, heat
 and oil for agricultural machinery.
A detailed description of the implementation of each sector is provided in \cite{neumann_potential_2023}.

\noindent \textit{Germany-specific adjustments}

\noindent To increase spatial and sectoral detail for Germany, some sector-specific models are used instead of the general European datasets.
Transport demand and the distribution of drive technologies are adapted from the \textit{Aladin} model which is an agent-based simulation model that determines the market diffusion of drive systems \cite{plotz_modelling_2014}.
For 2050, the main drive technology in Germany is found to be electric (\SI{98.8}{\percent}).

The bottom-up simulation model \textit{Forecast} is used to determine the industrial transformation parameters \cite{noauthor_forecast_nodate}.
The latter provides projections for production volumes and primary energy shares in energy-intensive sectors such as steel and aluminium manufacturing.
We follow developments in demand and technology mix in line with a national study conducted to evaluate the energy transition \cite{luderer_energiewende_2025}.
For countries other than Germany, we rely on European datasets, including Eurostat \cite{eurostat_complete_2022} and the JRC IDEES database \cite{jrc}.

\textit{PyPSA-DE} co-optimizes the expansion and operation of capacities and infrastructure to meet all energy and feedstock demand in a carbon-neutral energy system.
Investment and operating parameters for the year 2050 are primarily taken from the \textit{technology-data} database v0.10.1 \cite{technology-data}, with most cost assumptions derived from the Danish Energy Agency data and discounted at a rate of \SI{7}{\percent}.
A selection of technology parameters is listed in Appendix~\ref{A2}.

The model includes infrastructure for electricity, hydrogen and carbon dioxide transport.
The electricity grid is based on OpenStreetMap data and incorporates planned projects from the Ten Year Network Development Plan and the German national electricity plan (\textit{Netzentwicklungsplan}) \cite{xiong_modelling_2025}.
Hydrogen transport infrastructure can be developed either through new pipeline construction or by repurposing existing natural gas pipelines.
As fossil gas plays only a minor role in a carbon-neutral system, natural gas transport is not explicitly modelled but is assumed to be unconstrained across Europe.

For power generation, we consider renewable energy sources (hydro, photovoltaic, onshore and offshore wind) combined with dispatchable hydrogen and gas power plants.
Gas plants can be equipped with carbon capture units when their operational characteristics justify the additional investment.
The model is complemented with storage technologies that temporally decouple energy generation and feedstock supply from consumption.
Energy carriers (electricity, heat, hydrogen) and feedstock (methanol, Fischer-Tropsch, HBI, ammonia) can be stored with suitable technology.

\noindent\textit{Renewable Potential and Availability}

Land availability for renewable deployment is determined using the CORINE land use database together with the Natura 2000 dataset, which allows us to exclude protected areas and other unsuitable sites, such as coastal lagoons or estuaries for offshore wind \cite{european_environment_agency_corine_2019, european_environment_agency_natura_2024}.
To reflect soft factors such as social acceptance, we adopt modest capacity densities for the remaining land: \SI{3}{MW/km\textsuperscript{2}} for both onshore and offshore wind and \SI{5.1}{MW/km\textsuperscript{2}} for solar.
These values are intended to represent economically and socially feasible potentials rather than absolute technical limits.
Because solar is the lowest-cost electricity source, we conduct a sensitivity run with a higher capacity density of \SI{25}{MW/km\textsuperscript{2}} pushing toward the technical limit. The results are presented in Appendix~\ref{A4}.

Renewable availability time series are derived using \textit{atlite} \cite{Hofmann_atlite_A_Lightweight_2021,c3s_era5_2018}, based on historical weather data from 2019.
Since the spatial resolution in European countries except for Germany is coarse, we subdivide the renewable potentials in five subregions each.
A single aggregated region would smooth out spatial heterogeneity and miss areas with high renewable availability.
With five subregions, domestic variation is retained, allowing both high- and low-potential areas to be represented.

The \textit{atlite} data is also used to model heating demand, ensuring that periods of low renewable generation coinciding with high heating demand ('dark lulls') are adequately represented.
The optimization covers one full year with a 4-hourly temporal resolution.
To improve the representation of critical system states, a segmentation algorithm \cite{kotzur_tsam_nodate} refines the temporal resolution during extreme events such as low renewable availability or high demand while reducing it during periods with little variability.
This was found to retain a similar accuracy to hourly modelling while remaining computationally manageable \cite{neumann_potential_2023}.
We assume a scenario in which all dispatchable and renewable generation capacities from today reached their end of life but include existing infrastructure.

\newpage
\noindent\textit{Boundary Conditions}

We do not enforce expansion of renewable energies in Germany except for \SI{30}{GW} of offshore wind by 2030 which aligns with the national target \cite{agora_energiewende_meer-wind_2024}.
This offshore capacity is either already in operation, under construction or planned.
Apart from land-use constraints for renewable energies, the system is subject to two boundary conditions: achieving climate neutrality and adhering to a carbon sequestration limit of \SI{200}{Mt} per year across the entire European model scope.
Since \textit{PyPSA-DE} only accounts for CO\textsubscript{2} emissions while the political target includes all greenhouse gas emissions the boundary condition calls for net negative CO\textsubscript{2} emissions of \SI{-50.5}{Mt} in Germany and net zero in Europe.
This compensates for emissions of methane, nitrous oxide, hydrofluorocarbons, perfluorocarbons, sulfur hexafluoride, nitrogen trifluoride as well as land use change.
The low spatial resolution in Germany's neighboring countries aggregates all cross-border electricity connections from German regions to a single node per country.
This simplification neglects potential bottlenecks that may arise from the regional distribution of power plants or concentrated renewable generation interacting with domestic grid constraints.
Therefore, we impose a maximum net electricity import limit of \SI{125}{TWh} per year with a maximum power of \SI{30}{GW} to account for potential grid bottlenecks.
For comparison, Germany's maximum cross-border power flow was \SI{18.8}{GW} in 2023 and the maximum annual net import from 2015 to 2024 was \SI{52.0}{TWh} \cite{bundesnetzagentur_marktdaten_nodate}.
To reduce reliance on a single country for hydrogen supply, we cap interregional hydrogen pipeline expansion at \SI{20}{GW}.
This is equivalent to one 48-inch pipeline (\SI{13}{GW}) and one 36-inch pipeline (\SI{7}{GW}) \cite{wang_european_202}.
These new pipelines are in addition to retrofitted natural gas infrastructure.

\noindent\textit{Industry Supply and Demand}

\begin{figure}[htbp]
    \centering
    \includegraphics[width=1.0\textwidth]{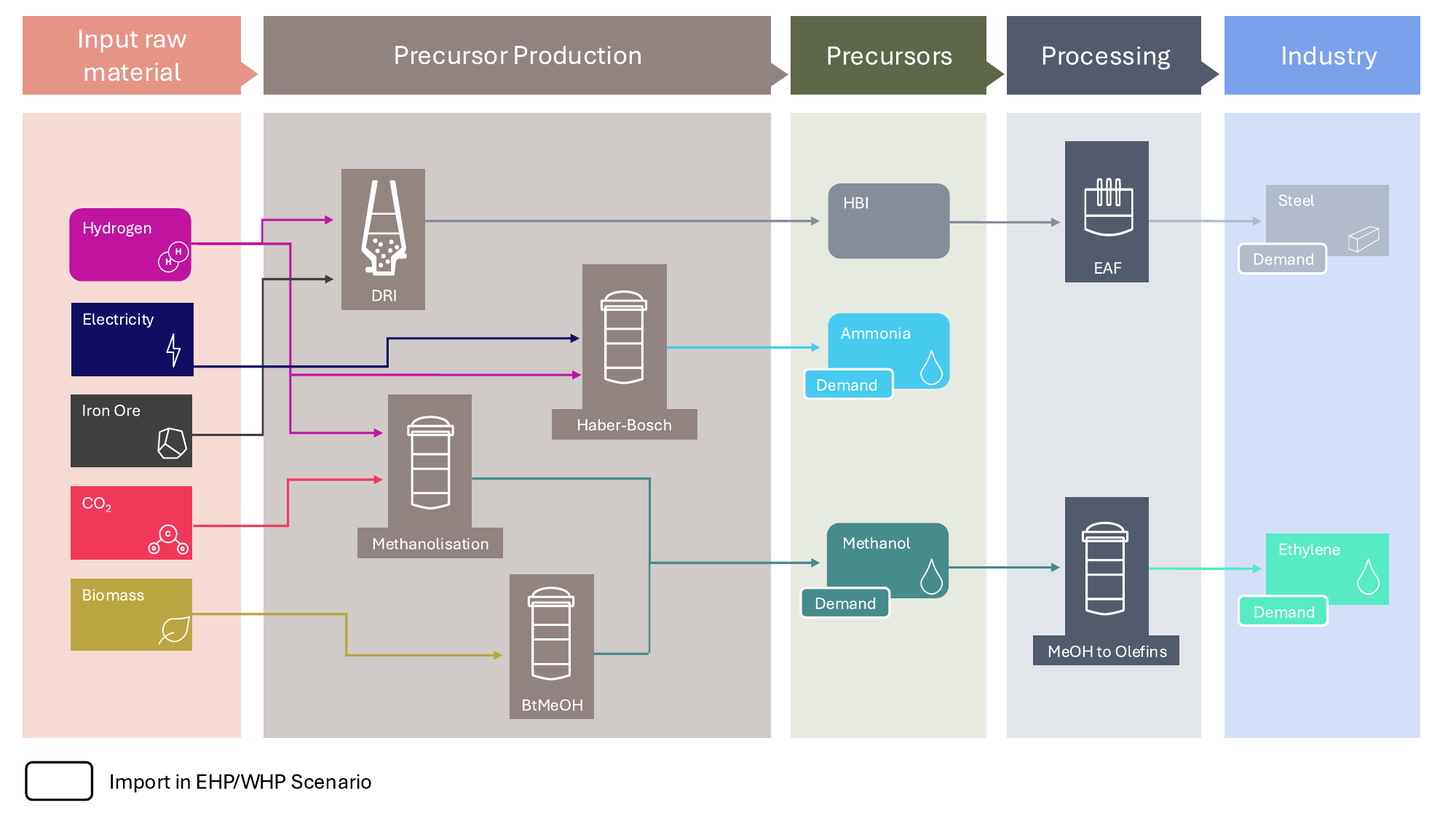}
    \caption{Endogenous industry representation.}
    \label{fig:industry_endogenized}
\end{figure}

To investigate the benefits of importing precursors, HBI, ammonia and methanol production is endogenized in \textit{PyPSA-DE} as depicted in Fig. \ref{fig:industry_endogenized}.
Instead of an exogenous demand, the investment in plants and their operation is added to the optimization problem.
For the production of steel only the primary route is modelled.
The secondary steel processed in an EAF is translated into an electricity demand as before.
Iron ore is first reduced in a hydrogen-fired DRI process using hydrogen to remove the oxygen of the ore.
The product, HBI, is upgraded in an electric arc furnace (EAF) to further reduce the HBI and remove impurities.
The EAF process stays in the current industry location in order to preserve the synergies between steel production and the manufacturing industry.
The production of ammonia is represented by only one step.
Hydrogen and electricity for an air separation unit are needed as input streams for the Haber-Bosch process.
Methanol can either be produced via a biogenic or hydrogen route. The synthesis gas needed for the methanolisation process can either be provided by hydrogen and CO\textsubscript{2} or through biogas.
The basic chemical for the production of high value chemicals (HVC - mainly plastics) is ethylene.
The methanol-to-olefins process is used for providing ethylene.
The route for naphtha being cracked into shorter hydrocarbons is not incorporated since yields of aromatics and olefins are lower and the process is harder to control.
Exhaust heat from the Fischer-Tropsch, Haber-Bosch and methanol synthesis is not integrated in a district heating system to isolate input procurement costs from the benefit of utilising potential alternative revenue streams from selling secondary outputs.
Integration of exhaust heat into a district heating system can be beneficial however it is questionable whether synthesis plants will be located in densely populated areas.
Since the spatial resolution of our model does not cover those challenges, exhaust heat is not integrated but in a future energy system, additional benefits may arise.

The flexibility of the precursor production processes influences the price. If the plant is built beyond the capacity necessary to cover the annual demand running at full load, flexibility allows the production to run in minimum part load or even shut down in times of high energy prices. Since the flexibility of green production routes is uncertain, we assume conservative minimum part load behavior of \SI{80}{\percent} for EAF, \SI{90}{\percent} for DRI, \SI{30}{\percent} for methanolisation and the Haber-Bosch process as well as \SI{50}{\percent} for the Fischer-Tropsch synthesis.
Shutdowns of precursor production is not allowed.
For transporting HBI, ammonia and methanol from European partners to Germany a distance of \SI{2000}{km} must be covered by electric truck (cargo load of 25 t, consumption of \SI{110}{kWh/100km}) considering an average electricity price of \SI{62}{EUR/MWh} from model results leads to costs of \SI{5.5}{EUR/t} \cite{liimatainen_potential_2019}.

\noindent\textit{Expanding Spatial Scope Beyond Europe}

\textit{PyPSA-DE}'s scope is limited to European countries only. Expanding the spatial scope to incorporate non-European partners is computationally expensive.
To mitigate this, we add the costs of imports from 51 countries using an open model of global energy supply chains, \textit{TRACE}, following Neumann et al. \cite{neumann_green_2025}.
\textit{TRACE} builds a stand alone greenfield energy system in each country.
After supplying the domestic demand the model expands capacities to produce energy carriers for export.
Since the certification of sustainable biomass as carbon source is difficult and there is a lack of data for sustainable biomass resources globally, the synthesis of carbonaceous fuels is only available via direct air capture.
For exporting energy carriers we investigate hydrogen, Fischer-Tropsch fuel, methanol, ammonia and HBI.
The export volume from each country is limited to \SI{500}{TWh/a} or in case of HBI \SI{238}{Mt} which is equivalent to the volume of hydrogen needed for reduction of iron ore.
Weather data and technology costs are harmonized across \textit{PyPSA-DE} and \textit{TRACE}.
After production, the model takes into account the energy needed to transport the energy carriers to Europe.
This is either captured by a loss factor (e.g. the evaporation of liquid hydrogen) or by having to provide additional methanol for the transport via ship.
Fig.~C.1 shows the costs for methanol from non-European countries \cite{hampp_import_2023}.

\backmatter

\bmhead{Supplementary information}

A dataset of the model results will be made available on zenodo after peer-review. The code to reproduce the experiments is available at https://github.com/toniseibold/de-import.

\bmhead{Acknowledgements}

T.S. gratefully acknowledges funding from the Kopernikus-Ariadne project by the Federal Ministry of Research, Technology and Space (Bundesministerium für Forschung, Technik und Raumfahrt, BMFTR). We thank Johannes Hampp, Alice DiBella, Michael Lindner, Julian Geis, Eva Herrmann and many
others for useful discussions. The responsibility for the contents lies with the authors.

\bmhead{Author's Contribution}

T.S.: Conceptualization - Data curation - Formal Analysis - Investigation - Methodology - Software - Validation - Visualization - Writing - original draft
F.N.: Investigation - Methodology - Software - Supervision - Validation
F.U.: Methodology - review \& editing
T.B.: Conceptualization - Formal Analysis - Funding acquisition - Investigation - Methodology - Project administration - Supervision - Validation - Writing - review \& editing

\section*{Declarations}

The authors declare no competing interests.

\bmhead{Declaration of generative AI and AI-assisted technologies in the writing process}

During the preparation of this work the author(s) used ChatGPT in order to improve wording.
After using this tool/service, the author(s) reviewed and edited the content as needed and take(s) full
responsibility for the content of the published article.

\begin{appendix}
\renewcommand{\thefigure}{\thesection.\arabic{figure}}
\renewcommand{\appendixname}{Appendix}

\section{Optimization Problem Formulation} \label{A1}

The optimization problem is formulated as a linear programming problem with the objective of minimizing total annual system costs.
The cost function includes investment and operational expenditures for generation, storage, transmission and conversion infrastructure.

Investment costs are annualized using an annuity factor to reflect payments over the asset's lifetime.
The first four terms include annualized investment costs $c$, which depend on installed generator capacity $G$ for each technology $r$, storage capacity $E$ for technology $s$, transmission line capacity $P$ and conversion or transport capacity $F$.

The second part of the function sums operational costs $o$ across all time steps, weighted by $w_t$, which represents the number of hours in each time step. These weights add up to the total number of hours in a year, $\sum w_t = \SI{8760}{h}$. Operational costs include variable costs $o$ for generator dispatch $g_{i,r,t}$ and link dispatch $f_{k,t}$.

\begin{equation}
    \begin{split}
        \min_{G,E,P,F,g} 
        \biggr[
            \sum_{i,r} c_{i,r} \cdot G_{i,r} + 
            \sum_{i,s} c_{i,s} \cdot E_{i,s} +
            \sum_{l} c_l \cdot P_l + 
            \sum_{k} c_k \cdot F_k + \\
            \sum_{t} w_t \cdot 
            \biggl(
                \sum_{i,r} o_{i,r} \cdot g_{i,r,t} +
                \sum_{k} o_k \cdot f_{k,t}
            \biggl)
        \biggr]
    \end{split}
\end{equation}

Each bus $i$ represents both a geographic region and an energy carrier.
Our model includes carriers such as electricity, hydrogen and CO\textsubscript{2}, as well as industrial precursors HBI, methanol and ammonia.

The model includes several linear constraints. For example, renewable feed-in is limited by historical weather data.
The full set of constraints follows the formulation presented in Neumann et al. \cite{neumann_potential_2023}.

We focus on the equality constraints that ensure demand is met at every time step for each energy carrier and region.
We formulate the balance equation as

\begin{equation}
        \sum_r g_{i,r,t} +
        \sum_s h_{i,s,t} +
        \sum_k L_{i,k,t} \cdot f_{k,t}
        \; = \; d_{i,t} \; \; \leftrightarrow \; \lambda_{i,t} \; \; \forall i,t
\end{equation}

Here, $h$ represents storage charging or discharging and $L_{i,k,t}$ is the lossy incidence matrix.
This matrix includes the efficiency of all processes $f$ that can meet the demand.
It is non-zero only when the link originates from node $i$.

The Lagrange multiplier $\lambda_{i,t}$ (also known as the KKT multiplier) correspond to the marginal price for carrier and bus $i$ at time $t$.
If the demand $d_{i,t}$ is increased slightly, the system adjusts to supply the additional amount.
The resulting change in the objective function reflects the marginal price.

We use this marginal price to calculate consumer costs, as introduced in the main section of this paper.
\newpage
\section{Technology Parameters} \label{A2}
\begin{longtable}{@{}lllll@{}}
    \caption{Excerpt from \textit{technology-data} of the technologies most relevant to this work \cite{technology-data} in EUR\textsubscript{2020}. Own assumption where source is missing.} \label{tab:technology-parameters} \\
    \toprule
    technology & parameter & value & unit & source \\
    \midrule
    Solar      & FOM        & 2.07  & \%/year      & \cite{dea_electricity_heating} \\
        & VOM        & 0.01  & EUR/MWh\textsubscript{el}  &  \\
        & overnight investment & 408.72 & EUR/kW\textsubscript{el}   & \cite{dea_electricity_heating} \\
        & lifetime   & 40.0    & years        & \cite{dea_electricity_heating} \\
    Onshore wind & FOM        & 1.18   & \%/year     & \cite{dea_electricity_heating} \\
        & VOM        & 1.29   & EUR/MWh     & \cite{dea_electricity_heating} \\
        & overnight investment & 1141.43 & EUR/kW     & \cite{dea_electricity_heating} \\
        & lifetime   & 30.0     & years       & \cite{dea_electricity_heating} \\
    Offshore wind & FOM        & 2.17   & \%/year     & \cite{dea_electricity_heating} \\
        & VOM        & 0.02   & EUR/MWh\textsubscript{el} & \cite{dea_electricity_heating} \\
        & overnight investment & 1523.93 & EUR/kW\textsubscript{el} & \cite{dea_electricity_heating} \\
        & lifetime   & 30.0     & years       & \cite{dea_electricity_heating} \\
    Grid connection & FOM        & 2.0   & \%/year     &  \\
        & overnight investment & 148151 & EUR/kW\textsubscript{el} & \cite{dea_electricity_heating} \\
        & lifetime   & 40.0     & years       &  \\
    Electrolysis & FOM & 4.0 & \%/year & \cite{dea_renewable_fuels} \\
        & efficiency & 0.70 & per unit & \cite{dea_renewable_fuels} \\
        & efficiency-heat & 0.13 & per unit & \cite{dea_renewable_fuels} \\
        & overnight investment & 1000.0 & EUR/kW\textsubscript{el} & \cite{dea_renewable_fuels} \\
        & lifetime & 25.0 & years & \cite{dea_renewable_fuels} \\
    DAC & FOM & 4.95 & \%/year & \cite{dea_carbon} \\
        & compression-electricity-input & 0.15 & MWh/t\textsubscript{CO\textsubscript{2}} & \cite{dea_carbon} \\
        & compression-heat-output & 0.2 & MWh/t\textsubscript{CO\textsubscript{2}} & \cite{dea_carbon} \\
        & electricity-input & 0.4 & MWh\textsubscript{el}/t\textsubscript{CO\textsubscript{2}} & \cite{beuttler_role_2019} \\
        & heat-input & 1.6 & MWh\textsubscript{th}/t\textsubscript{CO\textsubscript{2}} & \cite{beuttler_role_2019} \\
        & heat-output & 0.75 & MWh/t\textsubscript{CO\textsubscript{2}} & \cite{dea_carbon} \\
        & overnight investment & 4000000.0 & EUR/(t\textsubscript{CO\textsubscript{2}}/h) & \cite{dea_carbon} \\
        & lifetime & 20.0 & years & \cite{dea_carbon} \\
    Fischer-Tropsch & FOM & 3.0 & \%/year & \cite{agora_verkehrswende_future_2018}\\
        & VOM & 2.23 & EUR/MWh\textsubscript{FT} & \cite{dea_renewable_fuels} \\
        & capture rate & 0.9 & per unit & \cite{hannula_co-production_2015} \\
        & carbondioxide-input & 276 & t\_CO2/MWh\textsubscript{FT} & \cite{dea_renewable_fuels} \\
        & efficiency & 0.75 & per unit & \cite{dea_renewable_fuels} \\
        & electricity-input & 7 & MWh\textsubscript{el}/MWh\textsubscript{FT} & \cite{dea_renewable_fuels} \\
        & hydrogen-input & 1327 & MWh\textsubscript{H\textsubscript{2}}/MWh\textsubscript{FT} & \cite{dea_renewable_fuels} \\
        & overnight investment & 519739 & EUR/kw\textsubscript{FT} & \cite{agora_verkehrswende_future_2018} \\
    \botrule
    \newpage
    \toprule
    technology & parameter & value & unit & source \\
    \midrule
    Fischer-Tropsch & lifetime & 20.0 & years & \cite{dea_renewable_fuels} \\
    Methanolisation & FOM & 3.0 & \%/year & \cite{dea_renewable_fuels} \\
        & capture rate & 0.9 & per unit & \cite{hannula_co-production_2015} \\
        & carbondioxide-input & 248 & t\textsubscript{CO\textsubscript{2}}/MWh\textsubscript{MeOH} & \cite{bazzanella_technology_2017} \\
        & electricity-input & 271 & MWh\textsubscript{el}/MWh\textsubscript{MeOH} & \cite{bazzanella_technology_2017} \\
        & heat-output & 0.1 & MWh\textsubscript{th}/MWh\textsubscript{MeOH} & \cite{bazzanella_technology_2017} \\
        & hydrogen-input & 1138 & MWh\textsubscript{H\textsubscript{2}}/MWh\textsubscript{MeOH} & \cite{bazzanella_technology_2017} \\
        & overnight investment & 519739 & EUR/kW\textsubscript{MeOH} & \cite{agora_verkehrswende_future_2018} \\
        & lifetime & 20.0 & years & \cite{dea_renewable_fuels} \\
    Biomass-to-MeOH & C in fuel & 0.44 & per unit & \cite{millinger_are_2022} \\
        & C stored & 0.56 & per unit & \cite{millinger_are_2022} \\
        & CO2 stored & 0.21 & t\textsubscript{CO\textsubscript{2}}/MWh\textsubscript{th} & \cite{millinger_are_2022} \\
        & FOM & 2.67 & \%/year & \cite{dea_renewable_fuels} \\
        & VOM & 14.47 & EUR/MWh\textsubscript{MeOH} & \cite{dea_renewable_fuels} \\
        & capture rate & 0.9 & per unit & \cite{hannula_co-production_2015} \\
        & efficiency & 0.65 & per unit & \cite{dea_renewable_fuels} \\
        & efficiency-electricity & 0.02 & MWh\textsubscript{el}/MWh\textsubscript{th} & \cite{dea_renewable_fuels} \\
        & efficiency-heat & 0.22 & per unit & \cite{dea_renewable_fuels} \\
        & overnight investment & 1553.16 & EUR/kW\textsubscript{MeOH} & \cite{dea_renewable_fuels} \\
        & lifetime & 20.0 & years & \cite{dea_renewable_fuels} \\
    MeOHtO/A & FOM & 3.0 & \%/year & \cite{dea_renewable_fuels} \\
        & VOM & 31.75 & EUR/t\textsubscript{HVC} & \cite{bazzanella_technology_2017} \\
        & carbondioxide-output & 0.61 & t\textsubscript{CO\textsubscript{2}}/t\textsubscript{HVC} & \cite{bazzanella_technology_2017} \\
        & electricity-input & 1.39 & MWh\textsubscript{el}/t\textsubscript{HVC} & \cite{bazzanella_technology_2017} \\
        & overnight investment & 2781006.44 & EUR/(t\textsubscript{HVC}/h) & \cite{bazzanella_technology_2017} \\
        & lifetime & 30.0 & years & \cite{dea_renewable_fuels} \\
        & methanol-input & 18.03 & MWh\textsubscript{MeOH}/t\textsubscript{HVC} & \cite{bazzanella_technology_2017} \\
    DRI & FOM               & 11.3      & \%/year           & \cite{mpp} \\
        & electricity-input & 1.03      & MWh\textsubscript{el}/t\textsubscript{HBI}    & \cite{mpp} \\
        & hydrogen-input    & 2.1       & MWh\textsubscript{H\textsubscript{2}}/t\textsubscript{HBI}    & \cite{mpp} \\
        & overnight investment        & 4277858.0 & EUR/t\textsubscript{HBI}/h      & \cite{mpp} \\
        & lifetime          & 40.0      & years             & \cite{mpp} \\
        & ore-input         & 1.59      & t\textsubscript{ore}/t\textsubscript{HBI}     & \cite{mpp} \\
    EAF & FOM               & 30.0      & \%/year          & \cite{mpp} \\
        & electricity-input & 0.64    & MWh\textsubscript{el}/t\textsubscript{steel}  & \cite{mpp} \\
        & hbi-input         & 1.0       & t\textsubscript{HBI}/t\textsubscript{steel}   & \cite{mpp} \\
        & overnight investment        & 1839600.0 & EUR/t\textsubscript{steel}/h    & \cite{mpp} \\
        & lifetime          & 40.0      & years             & \cite{mpp} \\
    \botrule
    \newpage
    \toprule
    technology & parameter & value & unit & source \\
    \midrule
    Haber-Bosch & FOM              & 3.0      & \%/year         & \cite{dea_renewable_fuels} \\
        & VOM              & 0.02   & EUR/MWh\textsubscript{NH\textsubscript{3}}    & \cite{dea_renewable_fuels} \\
        & electricity-input& 0.25   & MWh\textsubscript{el}/MWh\textsubscript{NH\textsubscript{3}}& \cite{bazzanella_technology_2017} \\
        & hydrogen-input   & 1.15   & MWh\textsubscript{H\textsubscript{2}}/MWh\textsubscript{NH\textsubscript{3}}& \cite{bazzanella_technology_2017} \\
        & overnight investment       & 915.4941 & EUR/kW\textsubscript{NH\textsubscript{3}}     & \cite{bazzanella_technology_2017} \\
        & lifetime         & 30.0     & years           & \cite{dea_renewable_fuels} \\
        & nitrogen-input   & 0.16   & t\textsubscript{N\textsubscript{2}}/MWh\textsubscript{NH\textsubscript{3}}  & \cite{bazzanella_technology_2017} \\
    Air-separation & FOM              & 3.0      & \%/year         & \cite{dea_renewable_fuels} \\
        & electricity-input& 0.25   & MWh\textsubscript{el}/t\textsubscript{N\textsubscript{2}}& \cite{dea_renewable_fuels} \\
        & overnight investment       & 514601 & EUR/t\textsubscript{N\textsubscript{2}/h}     & \cite{dea_renewable_fuels} \\
        & lifetime         & 30.0     & years           & \cite{dea_renewable_fuels} \\
        & nitrogen-input   & 0.16   & t\textsubscript{N\textsubscript{2}}/MWh\textsubscript{NH\textsubscript{3}}  & \cite{dea_renewable_fuels} \\
    \botrule
\end{longtable}
\newpage
\section{Additional Model Results} \label{A3}

\textbf{Limitations}

\noindent The analysis is based on an optimization, investigating long-term system design.
This approach provides insights into the cost-optimal restructuring of the energy and industry sector but does not capture lock-in effects or transition dynamics.
It also limits the ability to assess the implications of Germany's 2045 climate neutrality target in relation to the EU's 2050 goal or to provide robust guidance on transition pathways.

Industrial precursor production routes, such as steel, are modeled in a simplified way.
While this abstraction omits route-specific dependencies (e.g. iron ore quality), it enables an initial assessment of how restructuring value chains affects the energy system.
The model also excludes soft factors, e.g. the availability of skilled labor and the tightly integrated nature of industrial clusters, where suppliers benefit from close proximity to product manufacturers.

The model exhibits a “the winner takes it all” behavior, where the most cost-effective option (e.g. the import of non-European HBI) is used exclusively, even if alternative options are only marginally more expensive.
This outcome reflects both the economic focus of the model and the absence of soft factors, such as geopolitical risks, political measures, supply chain resilience, or regional development considerations.

Our results are subject to considerable uncertainty, particularly with regard to the future deployment of electrolysis and carbon sequestration technologies and investment costs.
These factors could significantly influence the feasibility and attractiveness of different pathways, which is why we investigate the sequestration limit and investment costs of electrolysis units in Appendix~\ref{A4}.

\noindent \textbf{Comparing supply chain model TRACE with energy system model PyPSA-DE}

\begin{figure}[htbp]
    \centering
    \includegraphics[width=0.8\textwidth]{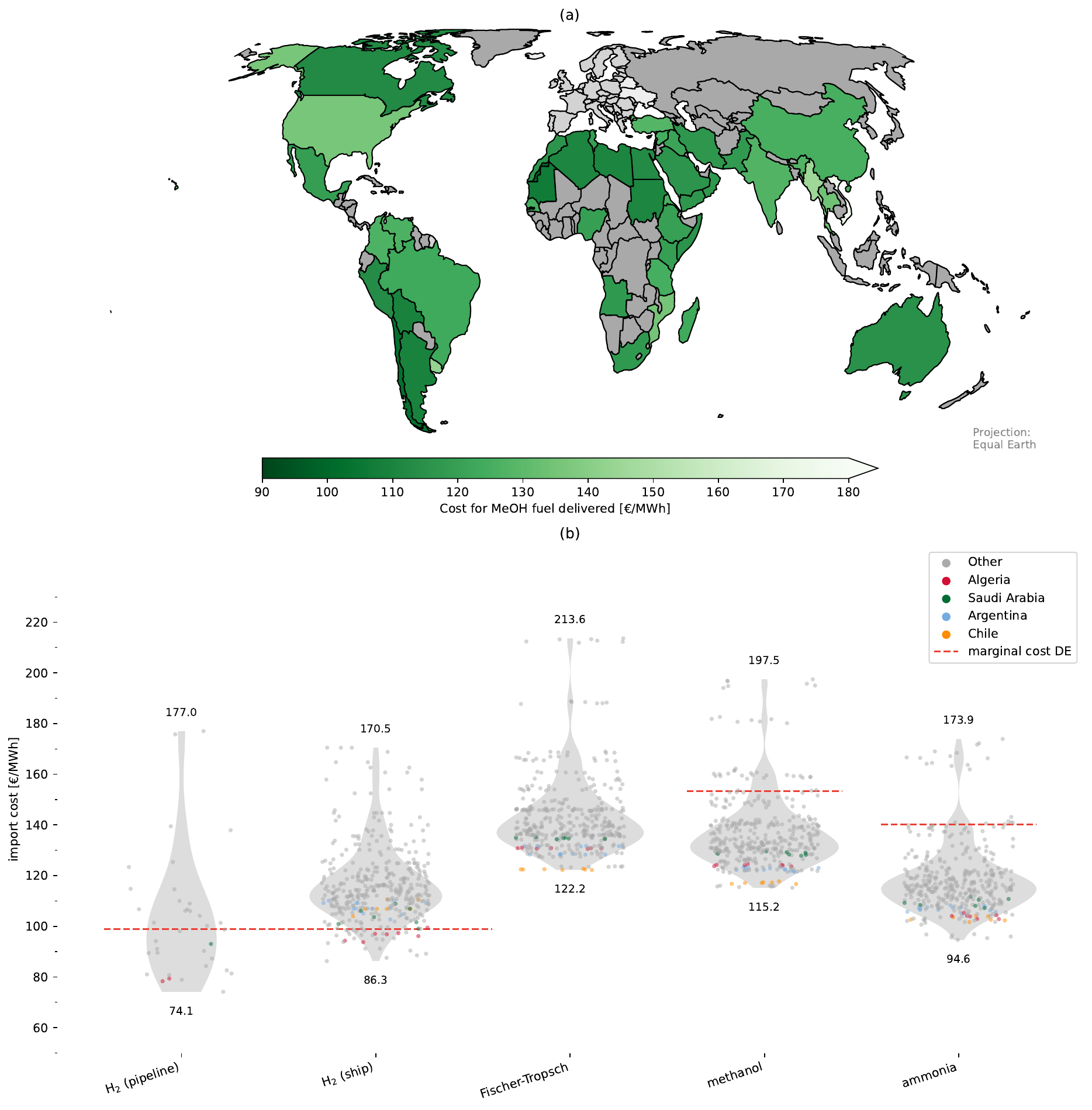}
    \caption{Import costs for methanol transported via ship from countries with high renewable energy potential to Europe following Neumann et al. \cite{neumann_green_2025} (a) and distribution of energy carriers prices compared to Base scenario prices in Germany (b). Fischer-Tropsch price in Germany is missing since the import is allowed across all scenarios which leads to an alignment of the price between Europe/Germany and the non-European countries.}
    \label{fig:trace_panel}
\end{figure}

\noindent \textit{TRACE} and \textit{PyPSA-DE} operate at different levels of detail and serve complementary purposes.
\textit{TRACE} is a supply chain model that focuses on the availability and costs of renewable energy carriers, while \textit{PyPSA-DE} is a much more detailed, sector-coupled energy system model that captures interactions across electricity, heat, mobility, industry and agriculture sector.
Fig.~\ref{fig:trace_panel} shows the costs of methanol delivered to Europe from varying non-European countries and compares commodity prices between \textit{TRACE} and \textit{PyPSA-DE}. 

To demonstrate the consistency of coupling these two models, we compare the levelized cost of electricity (LCoE) for two example countries:
Libya (as hydrogen exporter in the WH and WHP scenario) and Spain (best renewable resources in Europe) using \textit{TRACE}.
We then contrast the LCoE of Spain obtained from \textit{TRACE} with results from \textit{PyPSA-DE}.

\begin{figure}[htbp]
    \centering
    \includegraphics[width=0.8\textwidth]{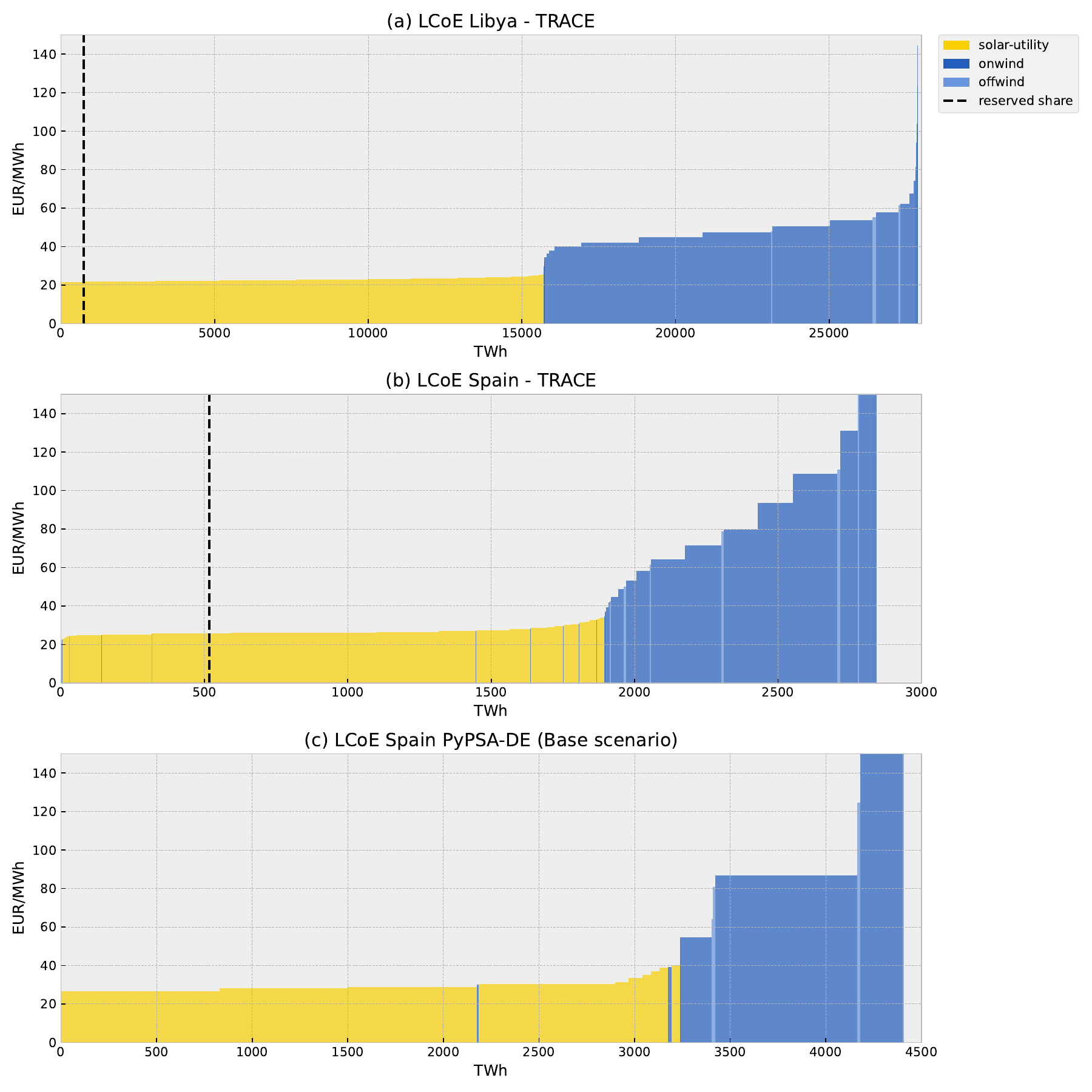}
    \caption{LCoEs for Libya (a) and Spain (b) using the supply chain model \textit{TRACE} in comparison with results from \textit{PyPSA-DE} under Base scenario assumptions (c).}
    \label{fig:lcoes}
\end{figure}

As depicted in Fig.~\ref{fig:lcoes}, the \textit{TRACE} results show that LCoE is lower in Northern Africa than in Spain, providing an economic rationale for hydrogen imports from the region (given the close link between hydrogen and electricity prices).
The Spain-Libya comparison illustrates several structural differences:
Spain has more limited renewable potential due to higher population density and mountainous terrain, while Libya (three times larger) offers abundant land and higher solar irradiation.
In both countries, solar power is the cheapest electricity source, but costs are lower in Libya.
This finding is consistent with our wider European hydrogen import studies, which identify Libya as one of the lowest-cost suppliers.

The LCoE for Spain derived from \textit{PyPSA-DE} under the Base scenario shows strong agreement with the \textit{TRACE} results.
While sector coupling in \textit{PyPSA-DE} introduces additional complexities such as electricity demand from heating, mobility, industry, and agriculture, and hydrogen exports from Spain to Central and Northern Europe the overall alignment confirms that integrating \textit{TRACE} with \textit{PyPSA-DE} is methodologically sound.

\begin{figure}[htbp]
    \centering
    \includegraphics[width=0.8\textwidth]{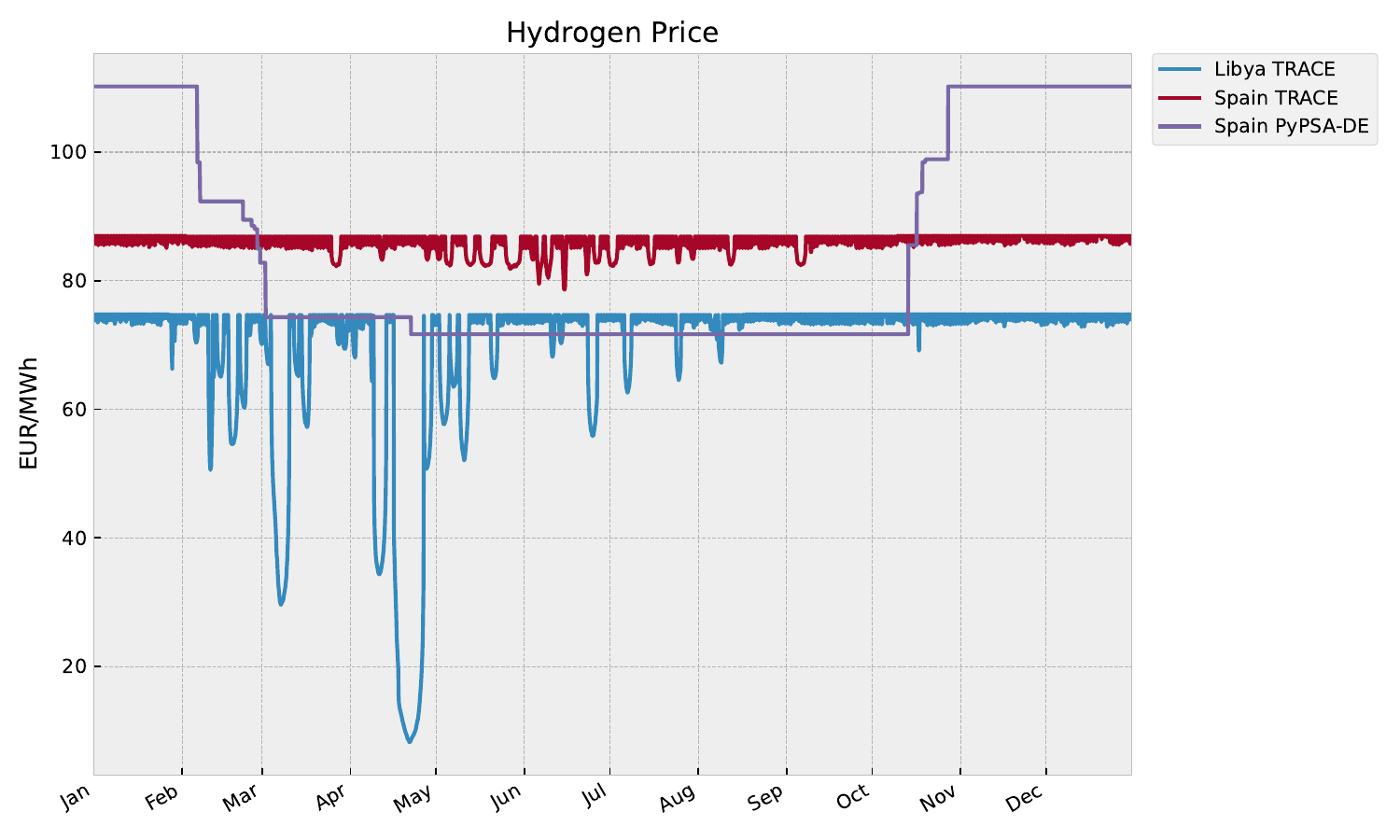}
    \caption{LCoHs for Libya (a) and Spain (b) using the supply chain model \textit{TRACE} in comparison with results from \textit{PyPSA-DE} under Base scenario assumptions (c).}
    \label{fig:lcohs}
\end{figure}

Another exercise to validate our approach is the comparison of hydrogen prices obtained from both models (Fig.~\ref{fig:lcohs}).
In \textit{TRACE}, the cheapest renewable resources are allocated to meet domestic electricity demand, but seasonal variations in load are not represented, nor are interactions with other sectors (notably heating), which increase system complexity.
With battery storage and hydrogen tanks included, this results in relatively stable hydrogen prices of \SI{69.98}{EUR/MWh} in Libya and \SI{86.01}{EUR/MWh} in Spain.

By contrast, \textit{PyPSA-DE} explicitly captures seasonal variability in hydrogen prices.
Spain also benefits from interconnections with neighboring grids and large-scale hydrogen storage in caverns, which help balance supply and demand.
Under these conditions, the hydrogen price averages \SI{79.75}{EUR/MWh} (demand weighted).

\noindent \textbf{Total consumer costs}

\begin{figure}[htbp]
    \centering
    \includegraphics[width=1.0\textwidth]{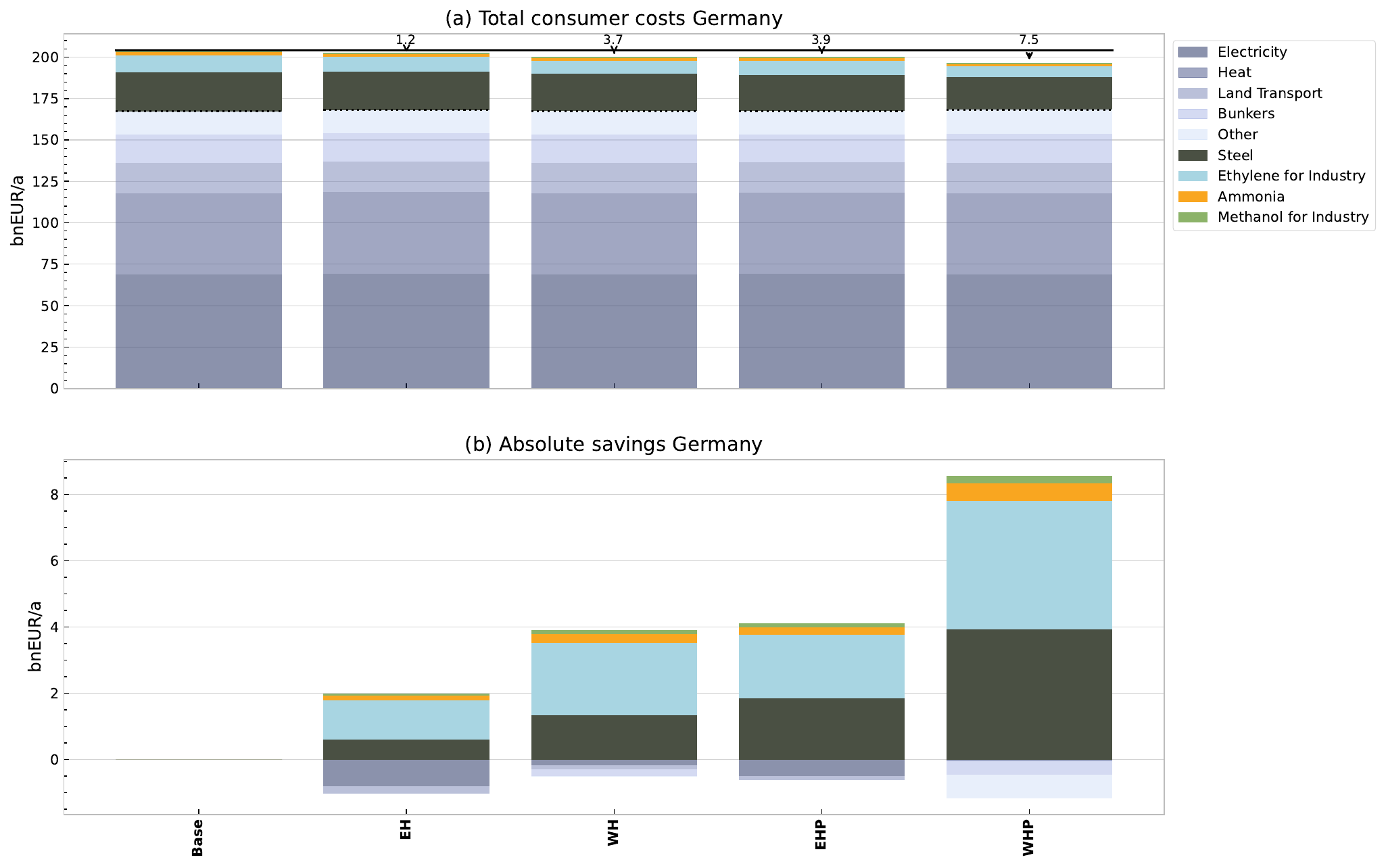}
    \caption{Total consumer costs (a) and absolute savings (b) in Germany}
    \label{fig:total_cc}
\end{figure}

\noindent The import options for hydrogen and precursors from various sources leads not only to declining industry consumer costs but also impacts the total consumer costs in Germany, that include all electricity, heating transport and industry demand as depicted in Fig.~\ref{fig:total_cc}.
In the Base scenario, consumer costs amount to \SI{204.0}{bnEUR/a}.
These costs decline as the model is given increasing flexibility in sourcing hydrogen and precursors.
The EH scenario delivers only modest savings of \SI{1.2}{bnEUR/a} (\SI{0.6}{\percent}).
Non-European hydrogen import leads to cost savings of \SI{3.7}{bnEUR/a} (\SI{1.8}{\percent}) and under the EHP scenario \SI{3.9}{bnEUR/a} (\SI{3.9}{\percent}).
The gains are even larger when global import options are considered, reaching \SI{7.5}{bnEUR/a} (\SI{3.7}{\percent}) in the WHP scenario.
Over all scenarios, the total consumer cost savings are lower than the industry savings.
This outcome reflects the model's objective of minimizing overall European system costs: slightly higher expenditures for electricity and heat in Germany are accepted if they benefit the wider European energy system.

\noindent \textbf{Hydrogen Balance}

\noindent In the Base scenario, Germany must meet its entire hydrogen demand domestically due to the boundary condition of zero hydrogen imports.
All demand is met via green electrolysis.
Due to the scarcity of renewable energy resources, hydrogen is primarily directed toward industrial applications, including high-temperature steam generation and the production of steel, ammonia and methanol.
Only a small share is repowered in backup capacities such as OCGT and CHP plants to cover dark lulls.
In contrast, the scenarios EH, WH, EHP and WHP introduce hydrogen imports, which fundamentally change the system's architecture.
In scenarios EH and WH, hydrogen imports from (non-)European partners reach 
\SI{188.1}{TWh} and \SI{288.8}{TWh}, respectively.
While the industrial hydrogen demand remains constant, the increased availability of imported hydrogen enables additional methanol production and a higher share of hydrogen in backup capacities especially in the WH scenario.
The EHP scenario Germany imports only \SI{13.4}{TWh}, as precursor production is relocated to other European countries and electrolysis is used to balance the variability of renewable energies.
In the WHP scenario, hydrogen imports rise to \SI{85.1}{TWh}, reflecting a system that has outsourced hydrogen-intensive processes like iron ore reduction to non-European partners.
The abundance of green imports allows Germany to use hydrogen for dispatchable capacities covering dark lulls.

\begin{figure}[htbp]
    \centering
    \includegraphics[width=0.8\textwidth]{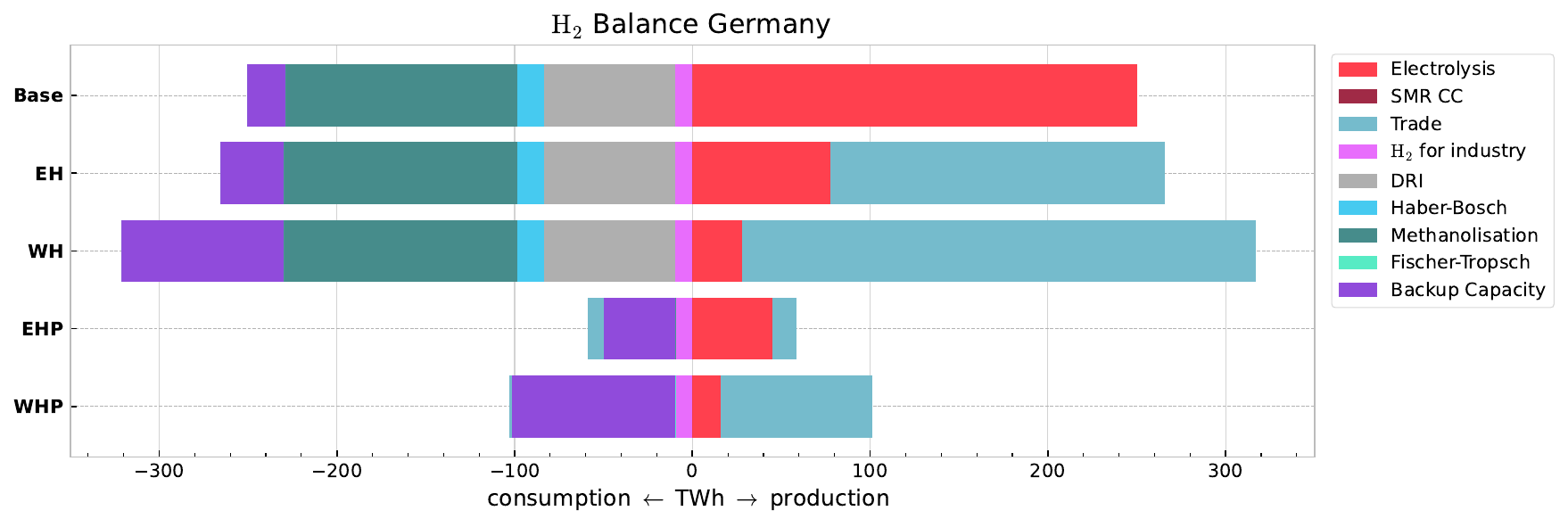}
    \caption{Germany hydrogen balance across scenarios.}\label{fig:hydrogenbalance}
\end{figure}

The picture for Europe follows the German trend.
In all scenarios, hydrogen is produced exclusively via electrolysis.
The Base scenario differs from the EH scenario only in the volume of produced hydrogen.
The surplus from European countries is provided for Germany.
Once industrial precursor production is allowed within Europe, hydrogen demand rises due to increased HBI, ammonia and methanol production that is shifted from Germany to European countries.
In the WHP scenario, access to a global market for these precursors reduces hydrogen demand in Europe (excluding Germany) to \SI{403.1}{TWh}.

\begin{figure}[htbp]
    \centering
    \includegraphics[width=0.8\textwidth]{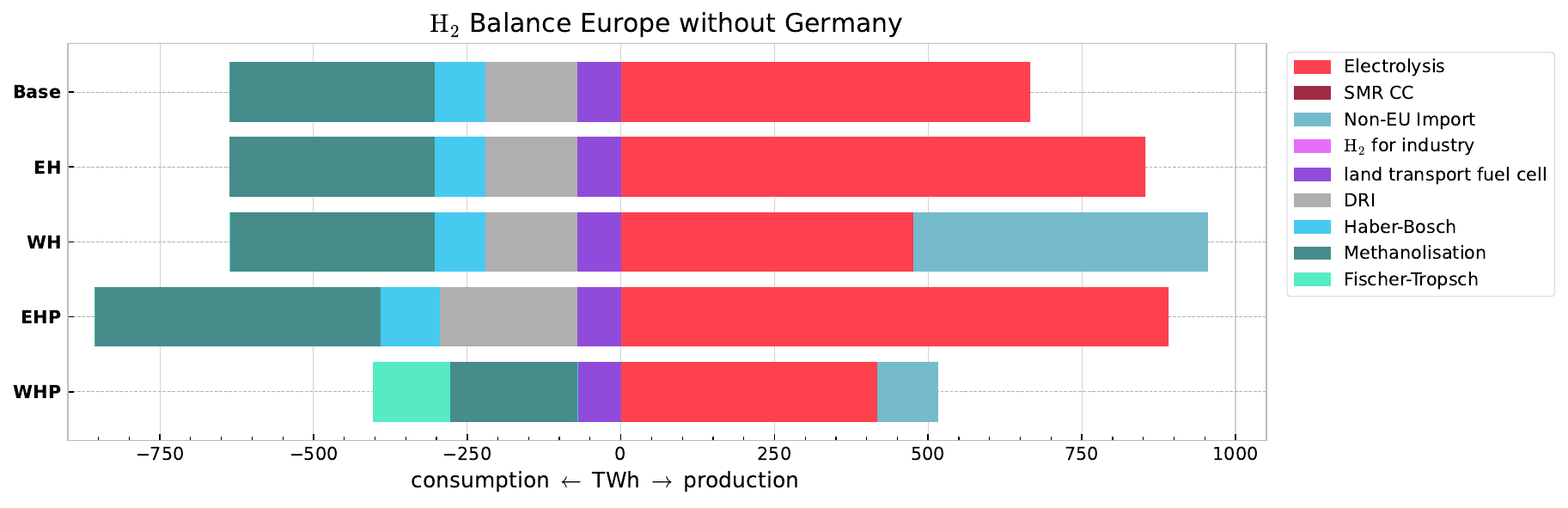}
    \caption{European (excluding Germany) hydrogen balance across scenarios. Higher production than consumption in scenarios EH, WH, EHP and WHP are provided to meet German hydrogen demand.}\label{fig:hydrogenbalance_eu}
\end{figure}
\newpage
\noindent \textbf{Carbon Balance}

\noindent Certain sources of CO\textsubscript{2} emissions remain constant across all scenarios, including emissions from industrial gas use, process-related emissions (e.g. from cement production), the use of solid biomass in industry and waste-based CHP plants.
These emission streams are difficult to eliminate entirely and form the baseline for Germany's carbon management needs.
Sequestration volume and CO\textsubscript{2} for methanolisation stays mostly constant as long as industry relocation is not allowed.
In the EHP and WHP scenario Germany's sequestration potential is utilized not only for domestic emissions but a rising volume is imported and sequestered.
In all scenarios the maximum of \SI{200}{Mt CO\textsubscript{2}} per year is sequestered in Europe.
The only difference is the location of emission and sequestration.

The system is designed to fulfill both the European and German climate target.
The German carbon price is represented by the sum of both shadow prices.
The carbon price varies across scenarios between \SI{320.29}{EUR/t CO\textsubscript{2}} and \SI{303.94}{EUR/ CO\textsubscript{2}}.
In general, CO\textsubscript{2} prices are lower than in other energy system optimizations since we allow the import of green fuels for the transport sector in all scenarios \cite{neumann_potential_2023}.

\begin{figure}[htbp]
    \centering
    \includegraphics[width=0.8\textwidth]{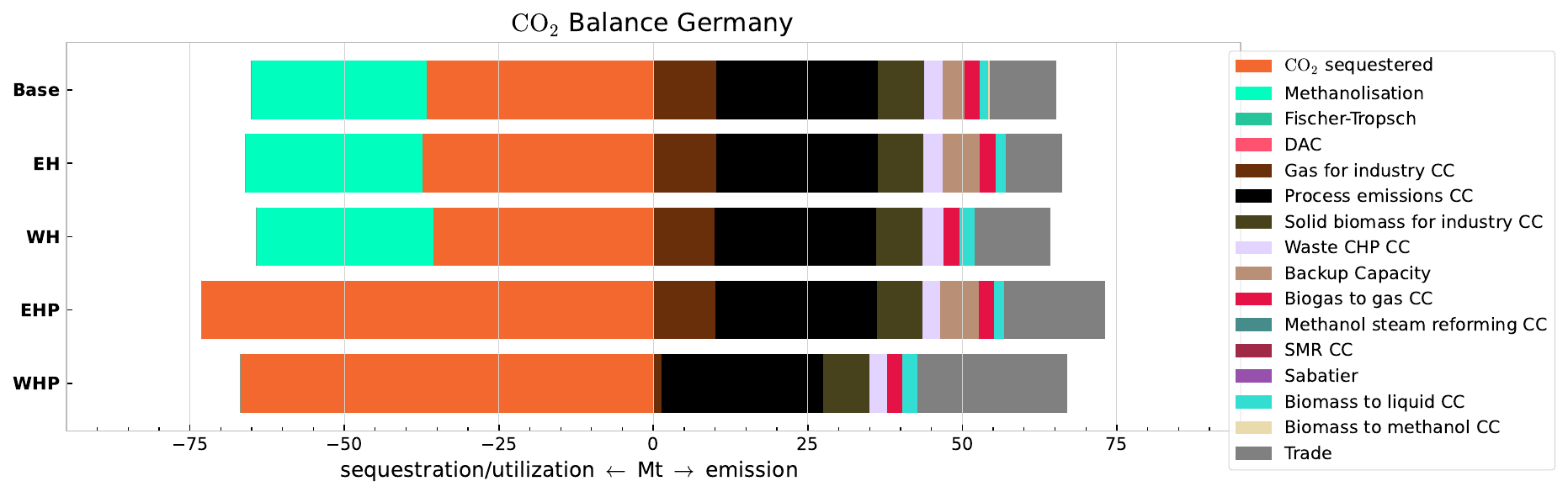}
    \caption{Germany CO\textsubscript{2} balance across scenarios.}\label{fig:co2balance}
\end{figure}

\noindent \textbf{Location of Precursor Production}

\noindent The availability of good renewable resources is the most influential factor in determining the location of industrial precursor production within the modeled scenarios.
In the Base, EH and WH scenarios, precursor production volumes and their locations remain fixed, as defined by the scenario design, regardless of regional cost differences.
However, once industrial relocation is introduced in the EHP scenario, production begins to shift toward regions with the lowest hydrogen prices.
Spain consistently offers the most cost-effective hydrogen due to its superior renewable energy potential, making it the primary production site for HBI, ammonia and methanol in the EHP scenario.
In the WHP scenario precursor production is mostly relocated to non-European partners with only methanol synthesis staying in Spain.
\begin{figure}[htbp]
    \centering
    \includegraphics[width=0.9\textwidth]{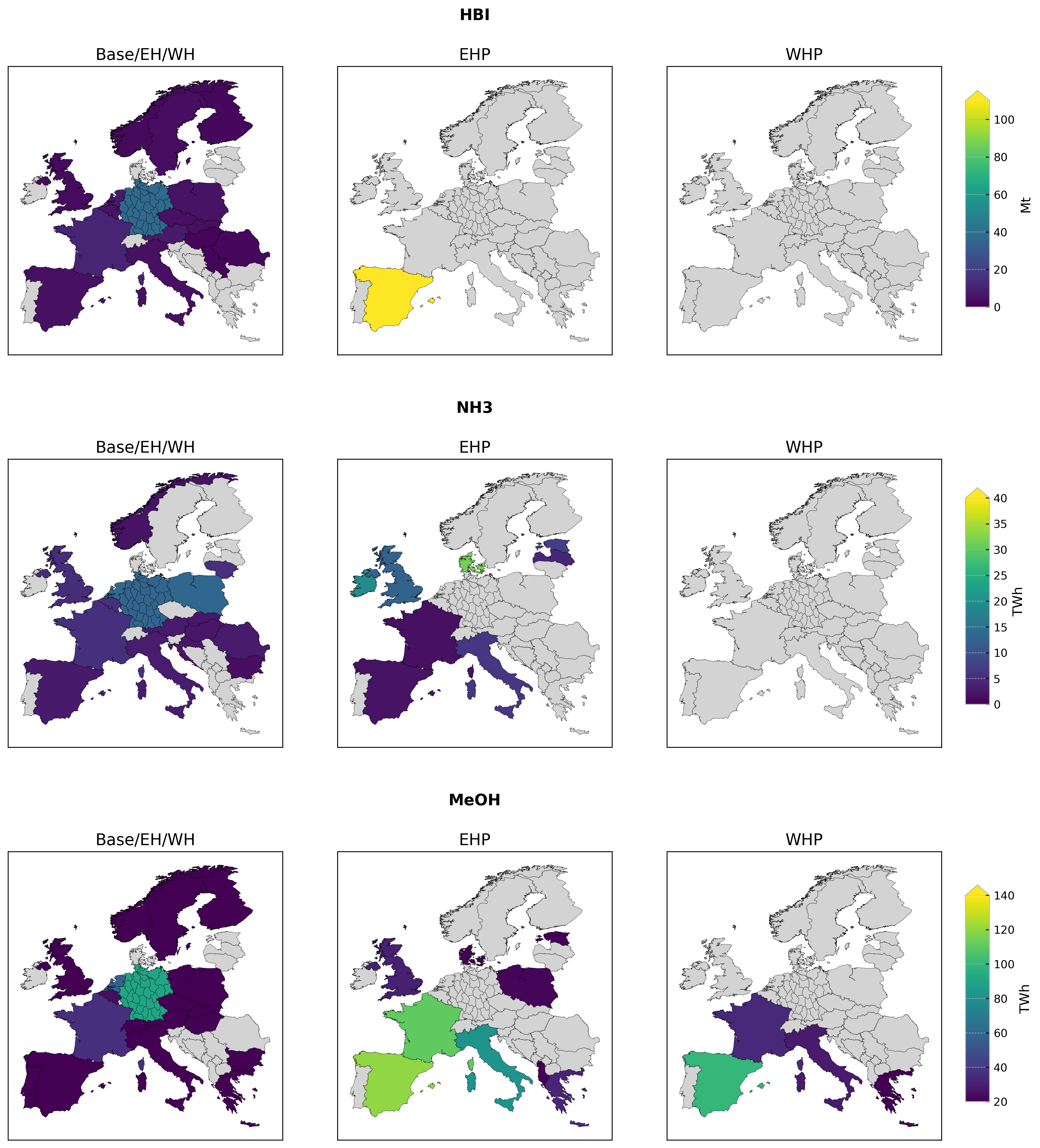}
    \caption{Production Volume of HBI, ammonia and methanol across scenarios. Locations and production volume is fixed for scenarios Base, EH and WH.}\label{fig:industrylocation}
\end{figure}

\section{Sensitivities}\label{A4}

The following section examines four sensitivity scenarios.
The first doubles the European CO\textsubscript{2} sequestration potential from 200 Mt CO\textsubscript{2} to 400 Mt CO\textsubscript{2} per year (\textbf{Seq}), while the second reduces capital expenditure for electrolysis units by \SI{50}{\percent} (\textbf{Elec50}) from \SI{1000}{EUR/kW} to \SI{500}{EUR/kW}.
A weighted average cost of capital (WACC) of \SI{7}{\percent} is applied in our default case, but depending on the country higher investment risk may increase financing costs.
To reflect this, we test a WACC of \SI{10}{\percent} for non-European countries (\textbf{wacc10}).
Finally, the solar potential is expanded from \SI{5.1}{MW/km\textsuperscript{2}} to \SI{25}{MW/km\textsuperscript{2}} to approximate technical limits (\textbf{HighSol}).
These parameters are uncertain, yet they exert a strong influence on system outcomes.

\begin{figure}[htbp]
    \centering
    \includegraphics[width=0.9\textwidth]{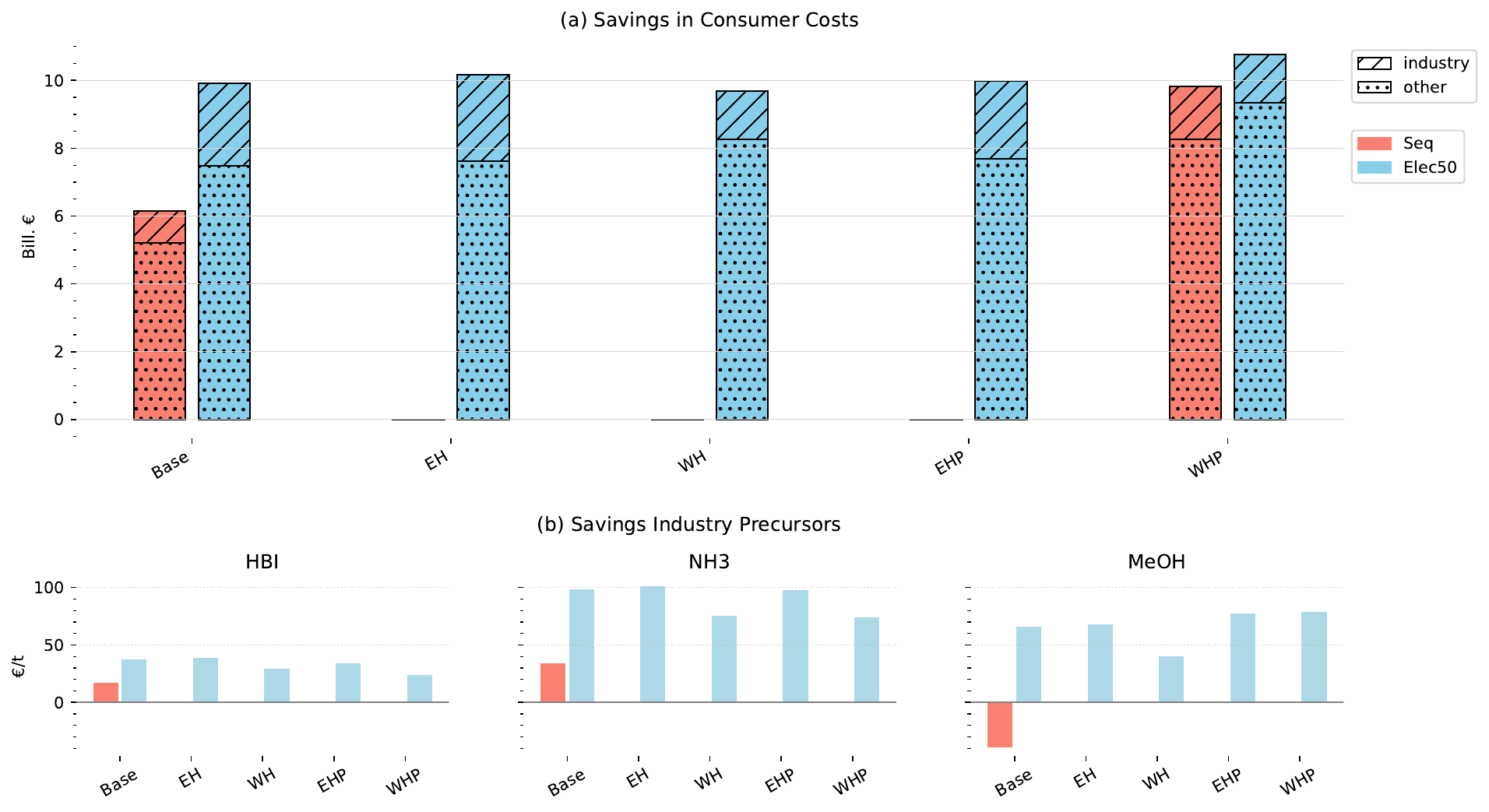}
    \caption{Cost differences compared to the scenarios presented in the main body of the paper. Consumer cost differences (top) and differences in precursor prices (bottom) for the sequestration sensitivity in red and the electrolysis cost sensitivity in blue.}\label{fig:sensitivity_costs}
\end{figure}

\begin{figure}[htbp]
    \centering
    \includegraphics[width=0.9\textwidth]{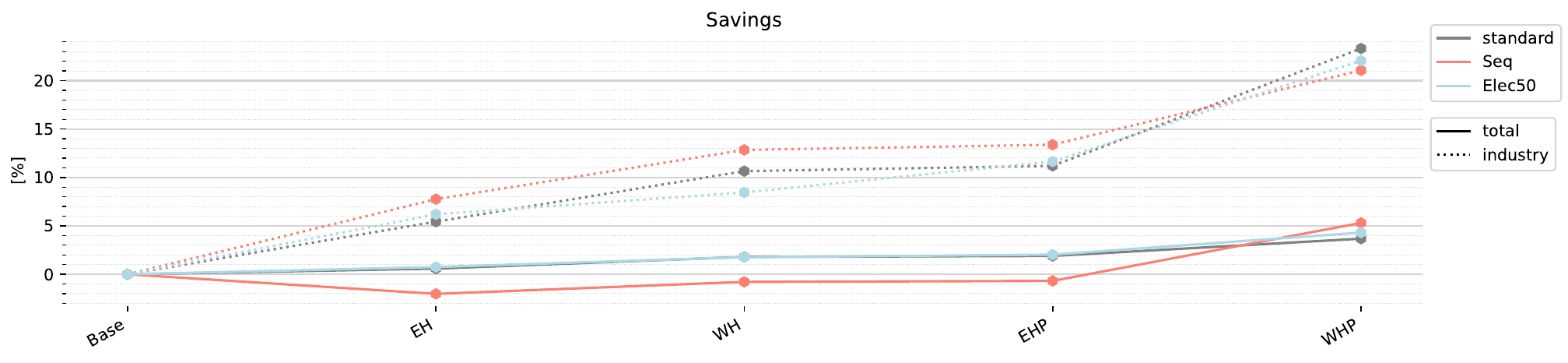}
    \caption{Cost savings overall and in the industry sector across the scenarios compared to each Base scenario.}\label{fig:sensitivity_savings}
\end{figure}

For the Seq sensitivity consumer costs are only decreasing in the Base and WHP scenario as depicted in Fig.~\ref{fig:sensitivity_costs}.
The higher sequestration potential allows Germany to use more CDR in order to offset emissions from fossil gas in the Base scenario and fossil oil in the WHP scenario.
However, savings are not exceeding \SI{9.8}{bnEUR/a}.
In the Elec50 sensitivity case, consumer costs decrease significantly across all scenarios.
The same effects can be seen in marginal prices of industry precursors.
Overall, the industry and total savings stay in the same range of the results presented in the main part of this paper (Fig.~\ref{fig:sensitivity_savings}).
Import from non-European partners offers a systemic benefit that cannot be reached in Europe alone.

\begin{figure}[htbp]
    \centering
    \includegraphics[width=0.9\textwidth]{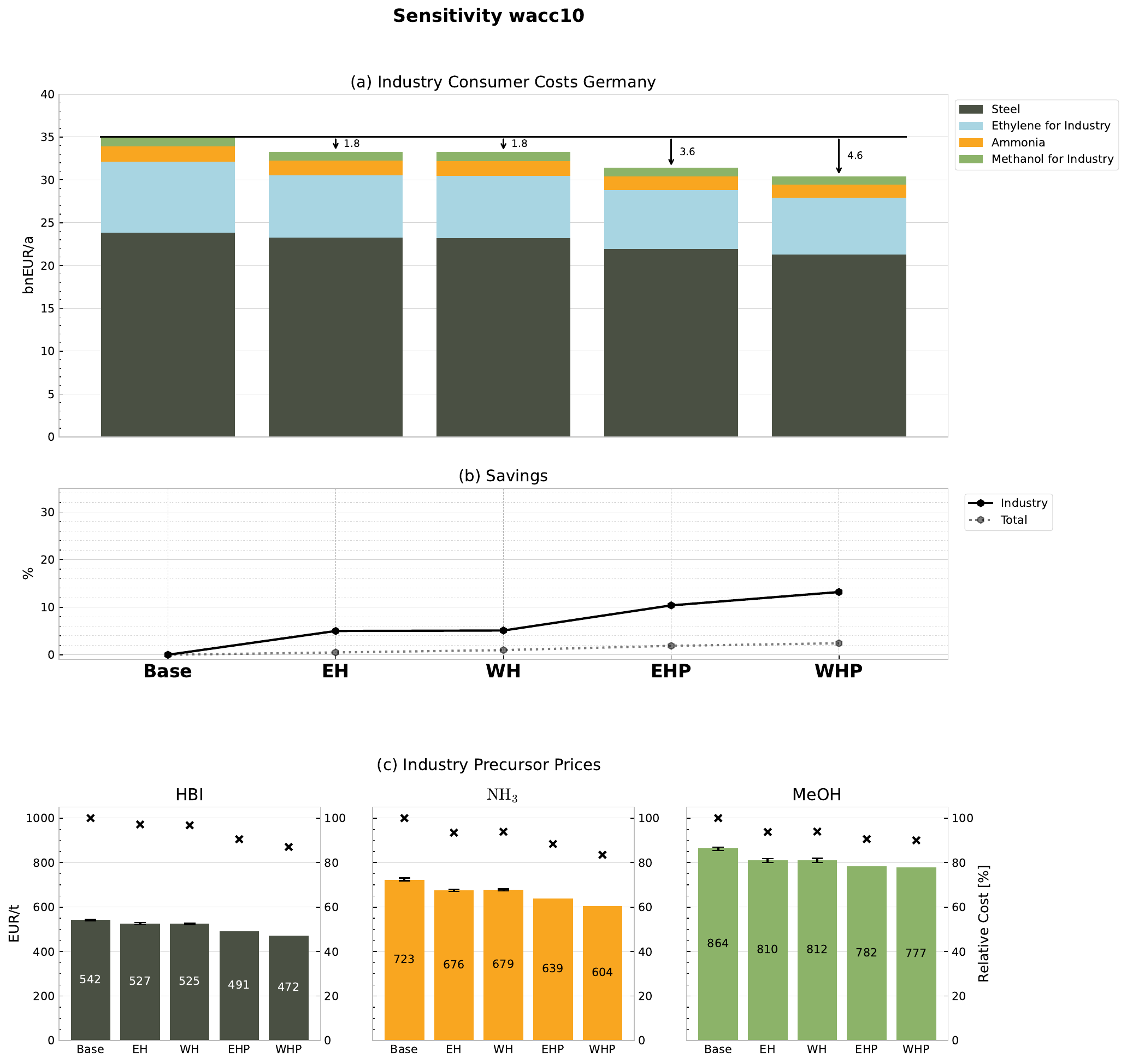}
    \caption{Industry consumer costs (a), relative savings (b) and industry precursor prices (c) in Germany under the assumption of a WACC of \SI{10}{\percent} in non-European countries.}\label{fig:cost_panel_wacc}
\end{figure}

The results paint a different picture when assuming a higher WACC in non-European countries.
As shown in Fig.~\ref{fig:cost_panel_wacc}, cooperating only with European countries becomes an attractive alternative to importing precursors from non-European countries. \SI{78.3}{\percent} of the WHP benefits can be reached.
However, total and industry consumer costs also increase significantly.

\begin{figure}[htbp]
    \centering
    \includegraphics[width=0.9\textwidth]{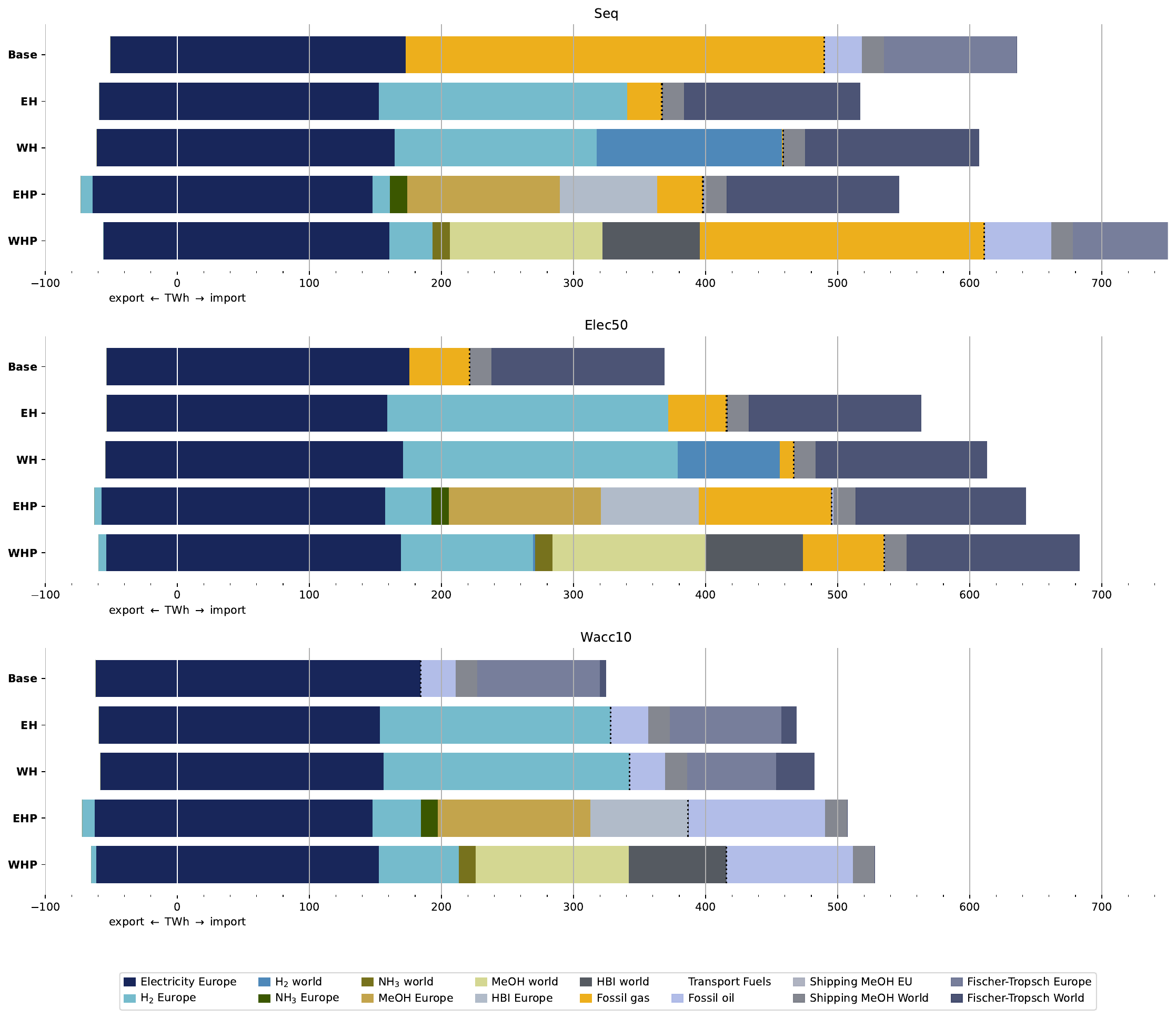}
    \caption{Import volumes in the sequestration (top), electrolysis (middle) and wacc (bottom) sensitivity cases.}\label{fig:sensitivity_import}
\end{figure}

As shown in Fig. \ref{fig:sensitivity_import}, the Seq case leads to an increased use of fossil fuel.
In particular, gas is primarily used in CHP plants equipped with carbon capture, supporting system reliability during dark lulls.
Additionally, fossil oil makes a significant contribution to meeting demand in the transportation sector in the Base and WHP scenario.

In all sensitivity cases, the production of ammonia, a significant share of methanol and the reduction of iron ore in the WHP scenario is consistently shifted to non-European partners.
Likewise, methanol imports for international shipping from non-European sources remain economically favorable across all scenarios.

Lower electrolysis investment costs lead to a reduction in fossil gas imports, improving overall system efficiency.
Conversely, higher sequestration capacity reduces dependence on non-European imports but increases reliance on fossil fuels which may also be imported.

\begin{figure}[htbp]
    \centering
    \includegraphics[width=0.9\textwidth]{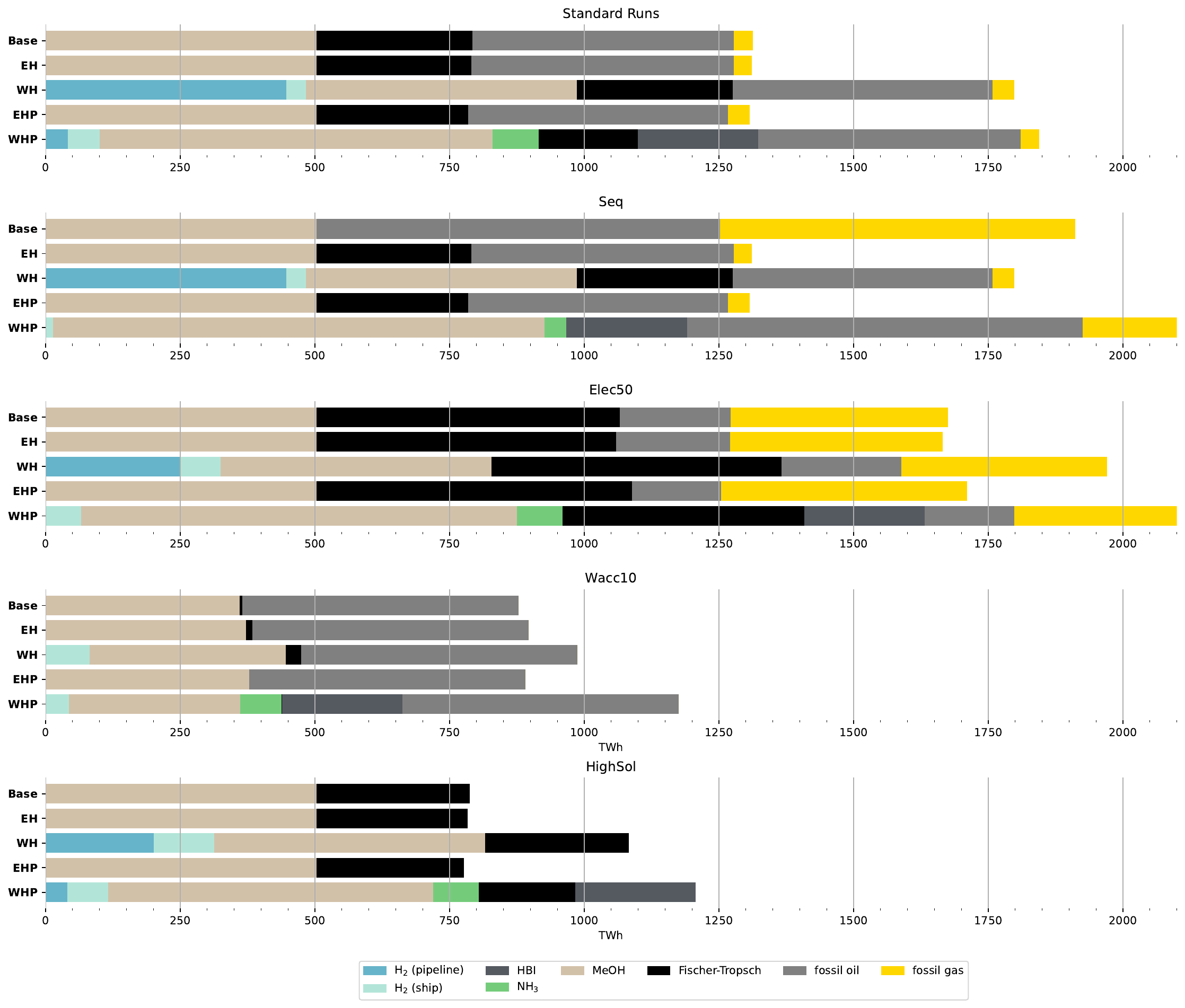}
    \caption{Import volumes from non-European partners to Europe in the standard run and the four sensitivity cases.}\label{fig:sensitivity_noneuimport}
\end{figure}

Fig.~\ref{fig:sensitivity_noneuimport} presents the total non-European import volumes to Europe.
Overall, import patterns remain largely consistent between the standard scenario and the sensitivity cases.
In case of the Seq sensitivity, Fischer-Tropsch imports are replaced by using fossil oil while the Elec50 sensitivity shows overall higher import volumes due to lower electrolysis investments which lead to lower import costs.
Increasing WACC leads to overall lower import volumes since the marginal prices of imports increase and become less attractive.
This also leads to higher CO\textsubscript{2} prices ranging from \SI{388.25}{EUR/t} to \SI{396.31}{EUR/t}.
This shows that climate targets are significantly harder to reach.

The HighSol sensitivity demonstrates how expanding technical limits reshapes Europe's energy system.
In the default case, the solar capacity cap for axis-tracking PV is binding in 38 regions, but with higher assumed potential this constraint remains binding in only 13 regions.
This relaxation enables greater exploitation of high-irradiation areas, confirming the central role of solar in Europe's decarbonization.
In the EHP scenario, total installed solar capacity rises from \SI[group-separator={,}, group-minimum-digits=4]{2897}{GW} to \SI[group-separator={,}, group-minimum-digits=4]{3031}{GW} with Spain showing the highest increase from \SI{456}{GW} to \SI{674}{SI}.

\begin{figure}[htbp]
    \centering
    \includegraphics[width=0.9\textwidth]{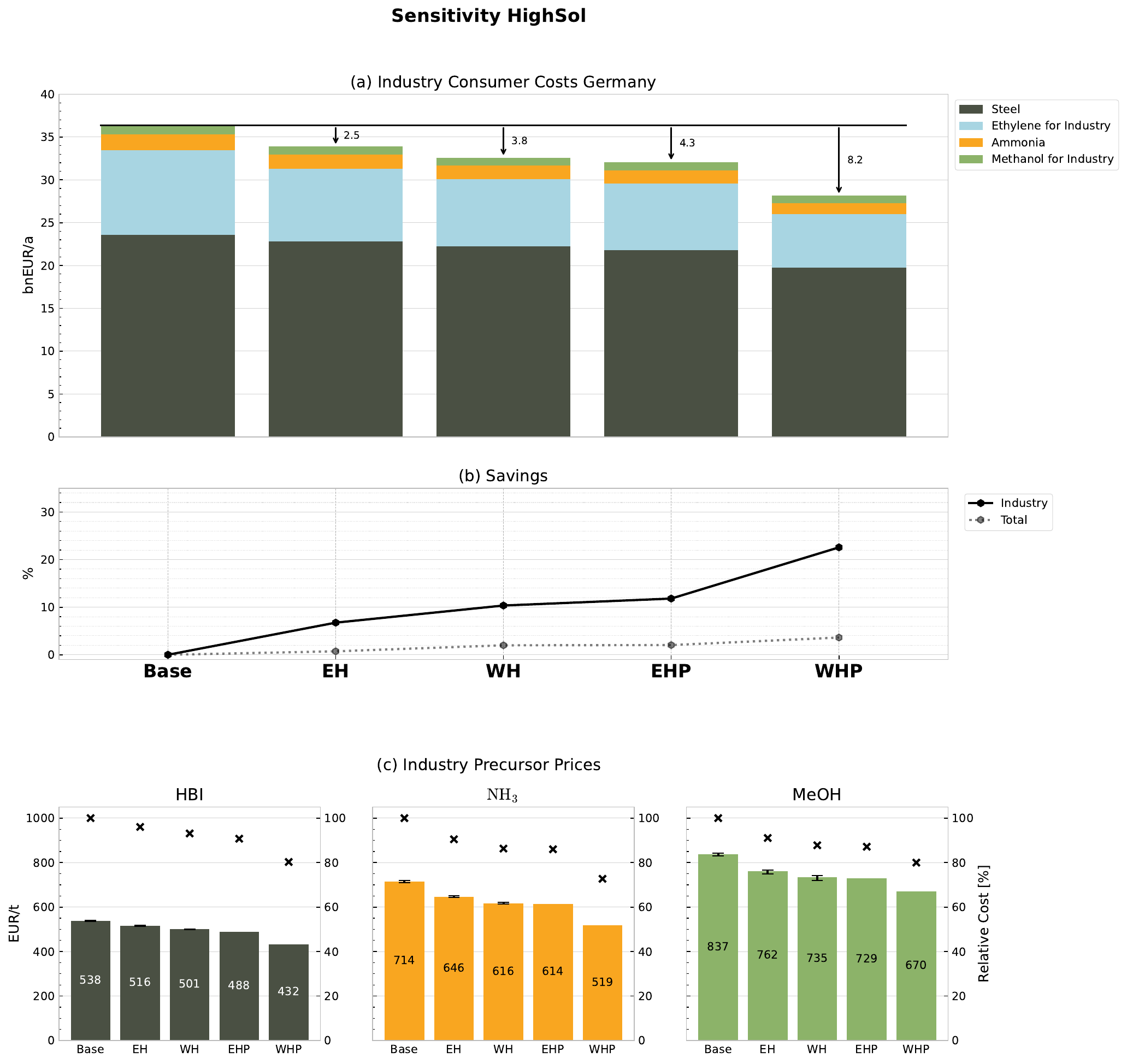}
    \caption{Industry consumer costs (a), relative savings (b) and costs for industry precursors (c) in Germany under the assumption of a higher availability of solar resources under the HighSol sensitivity run.}\label{fig:cost_panel_highsol}
\end{figure}

Fig.\ref{fig:sensitivity_noneuimport} shows that Europe becomes less dependent on non-European imports in the WH and WHP scenarios.
However, this reduction only partially translates into lower industry consumer costs for Germany, as shown in Fig.\ref{fig:cost_panel_highsol}.
In the Base scenario, industry consumer costs amount to \SI{36.3}{bnEUR/a}, which is lower than in the standard run presented in the main section of this paper.
As a result, the absolute cost savings in the WH and WHP scenarios are smaller.
Savings under the EH and EHP scenario are more pronounced due to the higher availabilty of solar energy.
Nevertheless, the EHP scenario achieves only \SI{52.4}{\percent} of the savings possible in the WHP scenario.
This underlines the unique advantage of non-European precursor imports, which cannot be matched by measures limited to the European context.

\end{appendix}
\clearpage
\bibliographystyle{apalike}
\bibliography{sn-bibliography}

\end{document}